\documentclass[11pt]{article}
\usepackage[english]{babel}
\usepackage[utf8]{inputenc}
\usepackage[T1]{fontenc}
\usepackage{bold-extra}
\usepackage{subfig}
\usepackage[subfigure]{tocloft}
\usepackage{comment}
\usepackage{amsfonts, amsmath, amsthm, amssymb}
\usepackage{amscd}
\usepackage{amsfonts}
\usepackage{anonchap}
\usepackage{mathtools}
\usepackage[hang, flushmargin]{footmisc}
\usepackage{footnotebackref}
\usepackage[table,xcdraw]{xcolor}
\usepackage{extsizes}
\usepackage[final]{graphicx}
\usepackage{float}
\usepackage{physics}
\usepackage{pdfpages}
\usepackage{slashed}
\usepackage[title,toc,titletoc]{appendix}
\usepackage{titlesec}
\usepackage{tocloft}
\usepackage{bbm}
\usepackage{tikz-cd}
\definecolor{rossoCP3}{cmyk}{0,.88,.77,.40}
\usetikzlibrary{decorations.pathmorphing}
\usepackage[sorting=none, style=nature]{biblatex}
\usepackage{ytableau}
\usepackage{youngtab}
\usepackage{listings}
\usepackage[margin=0.8in]{geometry}
\usepackage{xcolor}
\usepackage{footmisc}
\usepackage{footnotebackref}
\usepackage{hyperref}
\hypersetup{
    colorlinks,
    citecolor=black,
    filecolor=black,
    linkcolor=black,
    urlcolor=black
}

\titleformat{\section}
{\large\bfseries\scshape\centering}{\thesection\ }{1em}{}

\titleformat{\subsection}
  {\normalsize\bfseries\boldmath\centering}{\thesubsection}{1em}{}

	\definecolor{cadmiumorange}{rgb}{0.93, 0.53, 0.18}
  \definecolor{darkrosso}{RGB}{100,13,20}
\definecolor{lightor}{RGB}{248,150,30}
\newcommand{\ee}{\end{equation}}
\newcommand{\be}{\begin{equation}}
\newcommand{\bea}{\begin{align}}
\newcommand{\eea}{\end{alig}}

\newcommand   \cO {\mathcal{O}}

\newcommand{\identity}{\mathbbm{1}}

\title{\LARGE\color{rossoCP3}{\textbf{Charging the conformal window at nonzero $\theta$-angle}}}
\author{\normalsize Jahmall \textsc{Bersini}$^{\dagger\blacktriangle \color{rossoCP3}{\blacktriangle} }
$, Alessandra \textsc{D'Alise}$^{\star\color{rossoCP3}{\heartsuit}}$, Francesco \textsc{Sannino}$^{\ddagger \color{rossoCP3}{\diamondsuit}}$$^{\clubsuit\color{rossoCP3}{\heartsuit}}$$^\spadesuit$, Matías \textsc{Torres}$^{\sharp\color{rossoCP3}{\heartsuit}}$}

\date{}
\begin{document}
\maketitle 
{\let\thefootnote\relax\footnote{$^\dagger$ Electronic address: \textcolor{rossoCP3}{\href{mailto:jbersini@irb.hr}{\textcolor{rossoCP3}{jahmall.bersini@ipmu.jp}}}\\$^\star$ Electronic address: \textcolor{rossoCP3}{\href{mailto:alessandra.dalise@unina.it}{\textcolor{rossoCP3}{alessandra.dalise@unina.it}}}\\$^\ddagger$ Electronic address: \textcolor{rossoCP3}{\href{mailto:sannino@cp3.sdu.dk}{\textcolor{rossoCP3}{sannino@cp3.sdu.dk}}}\\$^\sharp$ Electronic address: \textcolor{rossoCP3}{\href{mailto:matiasignacio.torressandoval@unina.it}{\textcolor{rossoCP3}{matiasignacio.torressandoval@unina.it}}}}}\vspace{0.5mm} 
\begin{center}
\footnotesize{$^\blacktriangle$ Rudjer Boskovic Institute, Division of Theoretical Physics, Bijeni\v cka 54, 10000 Zagreb, Croatia }\\
\footnotesize{$^{\color{rossoCP3}{\blacktriangle}}$ Kavli IPMU (WPI), UTIAS, The University of Tokyo, Kashiwa, Chiba 277-8583, Japan}\\
\footnotesize{$\color{rossoCP3}{^\heartsuit}$ Dipartimento di Fisica ``E. Pancini", Università di Napoli Federico II - INFN sezione di Napoli, Complesso Universitario di Monte S. Angelo Edificio 6, via Cintia, 80126 Napoli, Italy}\\
\footnotesize{$^\clubsuit$ Scuola Superiore Meridionale, Largo S. Marcellino, 10, 80138 Napoli NA, Italy}\\
\footnotesize{$\color{rossoCP3}{^\diamondsuit}$ CP$^3$-Origins and D-IAS, Univ. of Southern Denmark, Campusvej 55, 5230 Odense M, Denmark}\\
\footnotesize{$^\spadesuit$ CERN, Theoretical Physics Department, 1211 Geneva 23, Switzerland} \\
\end{center}
\begin{abstract}
We determine the impact of the $\theta$-angle and axion physics on the near conformal dynamics of the large-charge baryon sector of $SU(2)$ gauge theories with $N_f$ fermions in the fundamental representation. We employ an effective approach featuring Goldstone and dilaton degrees of freedom augmented by the topological terms in the theory. We investigate how different dilaton potentials, including the ones for which a systematic counting scheme can be established, affects the results. Via state-operator correspondence we compute the corrections to the would-be conformal dimensions of the lowest large-charge operators as a function of the $\theta$ term and dilaton potential. \\
{\footnotesize  \it Preprint: RBI-ThPhys-2022-32}
\end{abstract}
\newpage
\tableofcontents

\section{Introduction}
\label{intro}

Understanding the non-perturbative dynamics of strongly coupled theories has proven a formidable challenge for theoretical physics. We concentrate here on gaining precious information on the near-conformal strongly coupled dynamics of two-color QCD for previously inaccessible sectors of the theory, i.e. the ones with large baryon charge and non-vanishing $\theta$-angle \cite{crewther1979chiral} including the presence of axions \cite{peccei1977cp,peccei1977constraints,Weinberg:1977ma,Wilczek:1977pj,Kim:1979if,Shifman:1979if,Dine:1981rt,Mohapatra:1978fy}. 
We do so by employing  and extending the formalism developed in \cite{Bersini:2022jhs,Orlando:2019skh,Orlando:2020yii}. Here, the near-conformal dynamics is enforced by introducing a dilaton state. This subject has received much attention over the years starting from the early work where the dilaton state was introduced at the effective action level  to saturate the underlying trace anomaly  for strongly coupled (supersymmetric) theories \cite{Schechter:1980ak,Rosenzweig:1979ay,Migdal:1982jp,Cornwall:1983zb,Salomone:1980sp,Ellis:1984jv,Gomm:1985ut,Veneziano:1982ah,Hsu:1998jd,Sannino:2002wb}. Recently attention has shifted to include the dilaton  \cite{Dietrich:2005jn, Park:2008zg, Chacko:2012sy, Matsuzaki:2013eva, Li:2016uzn, Kasai:2016ifi, Hansen:2016fri, Golterman:2016lsd, Golterman:2016cdd, Appelquist:2017wcg, Appelquist:2017vyy, Golterman:2018mfm, Brown:2019ipr, Appelquist:2019lgk, Cata:2019edh, Golterman:2020tdq, Golterman:2020utm, Appelquist:2020bqj, Golterman:2021ohm, Appelquist:2022mjb}  to  investigate near-conformal strongly coupled dynamics of theories close to the lower end of the conformal window for arbitrary matter representations \cite{Dietrich:2006cm,Sannino:2004qp}. For a summary of lattice results as well as relevant physical applications see \cite{Sannino:2009za,Cacciapaglia:2020kgq}.

For fixed charge physics  \cite{Hellerman:2015nra, Hellerman:2017sur, Alvarez-Gaume:2019biu, Orlando:2019hte, Badel:2019oxl, Antipin:2020abu, Gaume:2020bmp, Dondi:2022wli, Antipin:2022naw} the  dilaton (also referred to as the radial mode) was introduced \cite{Orlando:2020yii,Orlando:2019skh} in the absence of the topological $\theta$ term. The second ingredient is the interpretation via state-operator correspondence \cite{Cardy:1984rp,cardy1988universal} of the ground state energy on the cylinder as the lowest conformal dimension of the operators carrying nonzero baryon charge. We are, therefore, able to determine, for the first time, the dependence of  near-conformal scaling dimensions including the impact of the $\theta$ angle. 

A novelty  of the present work is the study of the impact of different dilaton potential models including the ones for which a systematic counting scheme can be established. 

 The paper is organized as follows. In  Section~\ref{dilaton} we introduce the dilaton for the two-color low energy effective theory at nonzero baryon chemical potential including the $\theta$-angle operator and axion field. Here we review the general dilaton theory emerging when deforming the underlying Conformal Field Theory (CFT) away from the fixed point. We summarize different dilaton potentials, investigated in the literature, encoding information on the conformal dimensions of the operators driving the CFT away from conformality.  Preparing for the state-operator correspondence we couple the theory to a nontrivial gravitational background with cylindrical topology. The classical vacuum structure on the cylinder is determined in Section~\ref{vacuum} in an expansion in inverse powers of the baryon charge \cite{Orlando:2020yii,}. This allows us to determine the $\theta$ dependence of the near-conformal scaling dimensions of the baryon charged operators, including the interplay on the quark and dilaton masses and potential. We further unveil a subtle dependence of the dynamics of the superfluid phase on the anomalous dimension of the fermion condensate. The corrections to the near-conformal scaling dimensions stemming from the quantum fluctuations arising from the spectrum of the theory are evaluated in Section~\ref{SOENC}. We offer our conclusions in Section~\ref{Conclusions}.

\section{Chiral Lagrangian near the conformal window}
 \label{dilaton}
 \subsection{Axion and $\theta$-angle Chiral Lagrangian}
Following \cite{Bersini:2022jhs}, the effective Lagrangian describing two-color QCD at finite baryon density is
\begin{align}
\label{lagtheta}
    \mathcal{L}= \nu^2 Tr\{ \partial_\mu\Sigma\partial^\mu\Sigma^\dagger\}+4\mu \nu^2 Tr\{B\Sigma^\dagger\partial_0\Sigma\}+m^2_\pi \nu^2 Tr\{M\Sigma+M^\dagger\Sigma^\dagger\} +2\mu^2 \nu^2\left[Tr\{\Sigma B^T\Sigma^\dagger B\}+Tr\{BB\}\right]  \, .
\end{align}
Here $\Sigma(t, x)$ is a matrix field that transforms in the two-index antisymmetric representation of $SU(2N_f)$, $4 \pi \nu$ is the energy scale of the chiral symmetry breaking and $\mu$ is the baryonic chemical potential. The mass $M$ and baryon charge $B$ matrices are given by
\begin{equation}
\label{Massa}
M =  \begin{pmatrix}
    0 & -\identity_{N_f}\\
    \identity_{N_f} & 0
    \end{pmatrix}  \,, \qquad B = 1/2\begin{pmatrix}
\identity_{N_f} & 0\\
0 & -\identity_{N_f}
\end{pmatrix}\ .
\end{equation}
In order to take into account the $\theta$-angle physics and the axial anomaly, the above needs to be augmented by the following term \cite{Kaiser:2000gs}
\be
\Delta\mathcal{L}_{\theta} =  -a \nu^2\left(\theta-\frac{i}{4}Tr\{\log \Sigma - \log \Sigma^\dagger \}\right)^2
\ee
while in the presence of the Peccei-Quinn axion field $N$ \cite{peccei1977cp,peccei1977constraints} one has to add
\be
\Delta\mathcal{L}_{\hat{a}}= \nu_{PQ}^2 \partial_\mu N\partial^\mu N^\dagger-a \nu^2\left(\theta-\frac{i}{4}Tr\{\log \Sigma - \log \Sigma^\dagger \}  -\frac{i}{4}a_{PQ}(\log N - \log N^\dagger) \right)^2\ ,
\ee
where $\sqrt{a}$ and $\sqrt{a_{PQ}}$ are the scales of the anomalous $U(1)_A$ and $U(1)_{PQ}$ symmetries, respectively, while $\nu_{PQ}$ is the scale of the spontaneous symmetry breaking of $U(1)_{PQ}$.

\subsection{Near-Conformal Chiral Lagrangian: The dilaton story}

To smoothly approach the conformal phase of the theory we non-linearly realize scale invariance by \emph{dressing} our Lagrangian via a dilaton field $\sigma (x)$, partially, serving as a conformal compensator \cite{coleman1988aspects,leung1986spontaneous,bardeen1986dilaton,yamawaki1986scale,Sannino:1999qe,Hong:2004td,Dietrich:2005jn,Appelquist:2010gy, goldberger2007light}. 

Therefore, under a scale transformation $x\mapsto e^{\alpha} x$, each operator $\mathcal{O}_k$ of dimension $k$ is assumed to transform as follows
\begin{equation}
    \mathcal{O}_k \mapsto e^{(k-4)\sigma f}\ \mathcal{O}_k\ ,
\end{equation}
where $f$ is the order parameter of the spontaneous scale symmetry breaking whose pseudo-Goldstone boson transforms as
\begin{equation}
    \sigma \mapsto \sigma-\frac{\alpha}{f}\ .
\end{equation}
 We  consider the fermion-induced mass term operator to have dimension $y=3-\gamma$, with $\gamma$ being the anomalous dimension of the fermion condensate constrained to be $0<\gamma<2$ by the unitarity bound. However, around $\gamma \simeq 1$ the underlying four fermion operator becomes near-marginal and therefore we will concentrate our analysis  in the interval  $0<\gamma<1$. 

However, in the absence of the underlying quark mass, the underlying conformal symmetry can break due to the emergence of another operator $\mathcal{O}$ with $\Delta_{\mathcal {O}}$ scaling dimensions just below the critical number of matter fields below which conformality is lost. This dynamics is encoded in the dilaton potential. This amounts to adding to the CFT the following Lagrangian term: 
\begin{equation}
\delta L_{\mathcal{O}} = \lambda_{\mathcal{O}} \mathcal{O} \ .
\end{equation}
 Here $\lambda_\mathcal{O}$ is the associated coupling to the Lagrangian. One expects this operator to be, for example, related to the emergence of a quasi-relevant four-fermion interaction \cite{Cohen:1988sq,Miransky:1996pd, Gies:2005as, Kaplan:2009kr, Fukano:2010yv,}. 
  
The general form of the dilaton potential is
\begin{equation}\label{sigmapot}
  \,V(\sigma) =   f^{-4} \,   e^{-4\sigma f} \sum_{n=0}^\infty\ c_n \, e^{-n(\Delta_\mathcal{O}-4)f \sigma }\ ,
\end{equation}
as can be shown by introducing a spurion field taking into account the explicit breaking of conformal symmetry \cite{rattazzi2001comments,goldberger2007light,chacko2013effective,appelquist2023dilaton}. Here $\Delta_\mathcal{O}$ is the scaling dimension of $\mathcal{O}$ while the $c_n$ depend on the given theory. Since we are interested in describing near-conformal dynamics we assume that  the explicit breaking is small. This can be realized when $\lambda_\mathcal{O} \ll 1$ and/or $\mathcal{O}$ is near marginal. In the former case one expects  $c_n \sim \lambda_\mathcal{O}^n$.

Inspired by the above ordering one could truncate the expansion \eqref{sigmapot} to the first two terms and obtain \cite{goldberger2007light}
\begin{equation}\label{sigmapotDelta}
  V(\sigma) =  \frac{m_{\sigma}^2 e^{-4 f \sigma } }{4 (4-\Delta_{\mathcal{O}} ) f^2}\left(1-\frac{4 }{\Delta_{\mathcal{O}} }e^{-(\Delta_{\mathcal{O}}-4)f \sigma }\right)  + \mathcal{O}( \lambda_{\mathcal{O}}^2) \ .
\end{equation}
Here the first two unknown coefficients have been fixed by requiring that the ground state occurs for $\sigma = 0$ and that the mass squared of $\sigma$ on the ground state  is $m^2_{\sigma}$. These constraints link the $c_0$ and $c_1$ coefficients as follows  
\begin{equation}
\frac{c_1}{c_0} = - \frac{4}{\Delta_{\mathcal{O}}} \ , \quad {\rm with} \quad c_0 = \frac{{f^2 \, m_{\sigma}^2   } }{4 (4-\Delta_{\mathcal{O}} ) } \ .
\end{equation} 
The coefficient of the cosmological constant $c_0$ is thereby forced to be of the same order as the first coefficient of the series  $c_1 \sim \lambda_{\mathcal{O}}$.  This  potential has been employed in recent investigations \cite{appelquist2023dilaton} including comparisons with lattice simulations \cite{appelquist2020dilaton}.   Assuming $\Delta_\cO = 2$ one recovers the usual $\phi^4$  Higgs-like potential while in the $\Delta_\cO \to 0$ limit one obtains
\begin{equation} \label{colman}
  V_{\Delta_\cO \to 0}(\sigma) =  -\frac{m_\sigma^2}{4 \Delta_\cO  f^2}-\frac{m_\sigma^2}{16 f^2}  \left(-4 f \sigma -e^{-4 f \sigma }+1\right) +\cO\left(\Delta_\cO \right)
\end{equation}
The order $\cO\left(\Delta_\cO^0 \right)$ term in the above expression coincides with the dilaton potential considered in the pioneering work of Coleman \cite{coleman1988aspects}.

Another interesting limit is the one for which the deformation itself is nearly marginal $\Delta_\mathcal{O} \to 4$. Here one expands eq.\eqref{sigmapot} in powers of $\Delta_\mathcal{O} - 4$ obtaining
\begin{equation} \label{logpot}
  V(\sigma) = -\frac{m_{\sigma}^2 e^{-4 f \sigma }}{16 f^2} (1+4 f \sigma ) +\mathcal{O}\left((\Delta_{\mathcal{O}}-4)^2\right)  \ .
\end{equation}
In fact, here one can  abide the conditions that the potential is minimised for $\sigma = 0$ and that at the  leading order in $\Delta_{\mathcal{O}} - 4 $ the mass for the dilaton is $m_\sigma$ without assuming an expansion in   $\lambda_{\mathcal{O}}$. The same potential can be derived from \eqref{sigmapotDelta} in the double  limit $\Delta_\mathcal{O} \to 4$ and $\lambda_{\mathcal{O}}\to 0$.  

The potential in \eqref{logpot} acquired central stage in a series of interesting papers by Golterman and Shamir \cite{Golterman:2016lsd, Golterman:2016cdd, Golterman:2018mfm, Brown:2019ipr, Golterman:2020tdq, Golterman:2020utm,  Golterman:2021ohm}. In these works the authors considered a dilaton EFT featuring the following counting in the  parameters ${p^2} \sim m \sim N_f-N_f^* \sim 1/N_c$.  $N_f^*$ is taken to be the critical number of fermions marking the onset of the IR fixed point, and $m$ is the quark mass. The identification between the potential in eq.(2)  of \cite{Golterman:2021ohm} and ours \eqref{logpot} occurs via the following map $f\sigma = -\tau$ and $m_\sigma^2/4 f^2 = f_\tau^2 B_\tau c^{GB}_1$.

In this work, we consider the  established potentials  for $\Delta_{\cO} \rightarrow 4$, $\Delta_\cO = 2$, and $\Delta_\cO \to 0$. In the latter case we disregard the divergent constant in eq.\eqref{colman}.

\subsection{Baryon charging the Dilaton-$\theta$-Axion-Chiral Lagrangian}
To access the (near) conformal dynamics of large charge operators we consider our system on a manifold $\mathcal{M}$ with volume $V$ and curvature $R$ such that the underlying new scale of the theory is $\Lambda_{Q}=(Q/V)^{1/3}$ where $Q$ is the fixed baryon charge. Concretely, we will take our manifold to be $\mathcal{M}= \mathbb{R}\times S^{d-1}$ such that we consider an approximate state-operator correspondence implying: 
\be
\Delta_Q =\tilde{V}^{1/3} E_Q\,, \qquad E_Q=\mu Q - \mathcal{L} \ ,
\ee
where $\Delta_Q$ is the scaling dimension of the lowest-lying operator with baryon charge $Q$, $E_Q$ is the ground state energy on $\mathbb{R}\times S^{d-1}$ at fixed charge, $\tilde{V}^{1/3}$ is the radius of $S^{d-1}$, and $\mu$ is the baryon chemical potential.

As customary, we introduce the chemical potential into the covariant derivative as the zero component of a gauge field. The dynamics of the theory is controlled by various energy scales. These are the chemical potential $\mu$, the mass of the quarks $m$, the scale of the axial anomaly $a$, the scales of chiral and conformal symmetry breaking $\nu$ and $f$, respectively, and the explicit conformal symmetry breaking scale $m_\sigma$. One can envision different counting schemes respecting the following hierarchy 
\begin{equation} \label{scales}
    m_\pi \,, m_\sigma  \ll \mu  \ll 4\pi \nu\ ,
\end{equation}
The first inequality implies that the theory is in the broken phase where pion condensation occurs \cite{Bersini:2022jhs} while the last inequality ensures the applicability of the chiral EFT at finite chemical potential.

After taking into account the background geometry and the dressing with the dilaton we arrive at the following two Lagrangians \cite{Monin:2016jmo,Son:2002zn,Hellerman:2015nra,coleman1988aspects,Orlando:2019skh}
\begin{equation}
\label{lagdressed}
    \begin{split}
    \mathcal{L}_{\theta, \sigma}&= \nu^2 Tr\{ \partial_\mu\Sigma\partial^\mu\Sigma^\dagger\}\ e^{-2\sigma f}+4\mu \nu^2 Tr\{B\Sigma^\dagger\partial_0\Sigma\}\ e^{-2\sigma f}+m^2_\pi \nu^2 Tr\{M\Sigma+M^\dagger\Sigma^\dagger\}\ e^{-y\sigma f}+\\& +2\mu^2 \nu^2\left[Tr\{\Sigma B^T\Sigma^\dagger B\}+Tr\{BB\}\right]\ e^{-2\sigma f} -a \nu^2\left(\theta-\frac{i}{4}Tr\{\log \Sigma - \log \Sigma^\dagger \}\right)^2\ e^{-4\sigma f} +\\&+\frac{1}{2}\left(\partial_\mu\sigma\partial^\mu\sigma-\frac{R}{6f^2}\right)\ e^{-2\sigma f}-V(\sigma)-\Lambda^4_0\ e^{-4\sigma f} \,,
\end{split}
\end{equation}
and
\begin{equation}
\label{lagaxion}
    \begin{split}
    \mathcal{L}_{\hat{a}, \sigma}&= \nu^2 Tr\{ \partial_\mu\Sigma\partial^\mu\Sigma^\dagger\}\ e^{-2\sigma f}+4\mu \nu^2 Tr\{B\Sigma^\dagger\partial_0\Sigma\}\ e^{-2\sigma f}+m^2_\pi \nu^2 Tr\{M\Sigma+M^\dagger\Sigma^\dagger\}\ e^{-y\sigma f}\\& +2\mu^2 \nu^2\left[Tr\{\Sigma B^T\Sigma^\dagger B\}+Tr\{BB\}\right]\ e^{-2\sigma f}+\nu_{PQ}^2 \partial_\mu N\partial^\mu N^\dagger\ e^{-2\sigma f}+\\& -a \nu^2\left(\theta-\frac{i}{4}Tr\{\log \Sigma - \log \Sigma^\dagger \}  -\frac{i}{4}a_{PQ}(\log N - \log N^\dagger) \right)^2\ \ e^{-4\sigma f}+\\&+\frac{1}{2}\left(\partial_\mu\sigma\partial^\mu\sigma-\frac{R}{6f^2}\right)\ e^{-2\sigma f}-V(\sigma)-\Lambda^4_0\ e^{-4\sigma f}\ ,
\end{split}
\end{equation}
where for later convenience we included the bare cosmological constant $\Lambda_0$.

\section{The vacuum structure and semiclassical expansion}
\label{vacuum}
In this section, we study the classical ground state energy of the theory which, according to the state-operator correspondence, gives the leading order in the large-charge expansion for the scaling dimension of the lowest-lying operator with baryon charge $Q$. As discussed in detail in \cite{Bersini:2022jhs} the ground state takes the following form 
\be
 \Sigma_0 = U(\alpha_i) \Sigma_c
\ee
with
\begin{align}
\label{eq:ansatsvacuum}
 U(\alpha_i) &= \text{diag}\{e^{-i \alpha_1}\,, \dots \,,e^{-i \alpha_{N_f}},e^{-i \alpha_1}\,, \dots \,,e^{-i \alpha_{N_f}} \} \qq{and} \nonumber \\ 
 \Sigma_c &=\begin{pmatrix}
    0 & \mathbbm{1}_{N_f}\\
    -\mathbbm{1}_{N_f} & 0
    \end{pmatrix}
\cos\varphi+i \begin{pmatrix}
    \mathcal{I} & 0\\
    0 & \mathcal{I}
    \end{pmatrix}\sin\varphi \qq{with} \mathcal{I}=\begin{pmatrix}
    0 & -\mathbbm{1}_{N_f/2}\\
    \mathbbm{1}_{N_f/2} & 0
    \end{pmatrix}\ ,
\end{align}
where the alignment angle $\varphi$ and the Witten variables $\alpha_i$ are determined by the equations of motion (EOMs).

Replacing this vacuum ansatz, the Lagrangian \eqref{lagdressed} becomes
\begin{equation}
\begin{split}
\small
    \mathcal{L}_{\theta, \sigma}\left[\Sigma_0,\sigma_0\right]&=-e^{-4 f \sigma_{0} } \Lambda_0 ^4-V(\sigma_0)-\frac{R\ e^{-2 f \sigma }}{12 f^2}+4 m^2_\pi \nu^2 X \cos\varphi\ e^{-f \sigma_{0}  y}\\&+2 \mu ^2 N_f \nu^2 e^{-2 f \sigma_{0} } \sin ^2\varphi -a \nu^2 e^{-4 f \sigma_{0} } {\bar \theta}^2\ ,
\end{split}
\end{equation}
where
\be
{\bar \theta} \equiv \theta -\sum_i^{N_f} \alpha_i \,, \qquad X \equiv \sum_i^{N_f} \cos \alpha_i  \ .
\ee
The respective equations of motion are
\begin{align}
N_f \mu ^2 e^{-2 f \sigma_0 } \cos \varphi -m_{\pi}^2 X e^{-f \sigma_0  y}&=0\\ \label{eqalphai}
a e^{-4 f \sigma_0 } \Bar{\theta}-2m_{\pi}^2 \sin \alpha_i \cos \varphi  e^{-f \sigma_0  y}&=0 \,, \qquad i=1, .., N_f \\
 \frac{R e^{-2 f \sigma_0 }}{6f}+ 4 a f \nu ^2 e^{-4 f \sigma_0 } Y^2+4 f \Lambda_{0}^4 e^{-4 f \sigma_0 }-\frac{\partial V(\sigma)}{\partial \sigma} {\bigg \rvert}_{\sigma=\sigma_0}+& \nonumber\\-4 f \mu ^2 N_{f} \nu^2 e^{-2 f \sigma_0 } \sin ^2\varphi 
    -4 f m_{\pi}^2  \nu^2 y X \cos \varphi  e^{-f \sigma_0  y}&=0\\
    4 \mu  N_{f} \nu^2 e^{-2 f \sigma_0 } \sin ^2\varphi &=\frac{Q}{V}\ .
\end{align}
We solve these equations in the large $Q$ expansion; focusing on the two extreme cases
$\gamma\ll 1$ and $1-\gamma\ll 1$, we find the following expressions for the ground state energy 

\begin{itemize}
    \item $\mathbf{V_{\Delta_\cO \to 0}(\sigma)}$
     \end{itemize}
\begin{equation}
\label{GSEnc}
\small
    \begin{split}
           E^{\gamma \ll 1}&=\ \frac{c_{4/3} Q^{4/3}}{\tilde{V}^{1/3}}+Q^{2/3}\tilde{V}^{1/3}\left\{c_{2/3}\tilde{R}-\frac{X_{00}^2}{4\pi^2 N_f^3 c_{4/3}^4}\bigg(\frac{9 m_\pi^2}{32\nu}\bigg)^2\left[1-\gamma\left(\frac{2}{3}\log Q -\log\left(\frac{32 N_f\nu^2 \pi^2c_{4/3} \tilde{V}^{2/3}}{3}\right) \right. \right. \right. \\& \left. \left. \left. -\frac{X_{10}}{X_{00}} \right) + \cO\left( \gamma^2\right)\right]\right\}-\tilde{V} \log Q \Bigg\{\frac{16\pi^2}{9}N_f c_{2/3}c_{4/3}\nu^2m_\sigma^2-\frac{\gamma}{3\pi^2 N_f^4 c_{4/3}^5}\bigg(\frac{9 m_\pi^2}{32\nu}\bigg)^2\Bigg[\frac{5}{8\pi^2 c_{4/3}^4 N_f^2}\bigg(\frac{9 m_\pi^2}{32\nu}\bigg)^2 X_{00}^4\\&-c_{2/3}\tilde{R} N_f X_{00}^2+\frac{9 X_{00} X_{01}}{32 c_{4/3}}\Bigg]  + \cO\left( \gamma^2\right)\Bigg\} +\cO\left(Q^0 \right)\ \qq{and}\\
     E^{1-\gamma\ll 1}&= \frac{c_{4/3} Q^{4/3}}{\tilde{V}^{1/3}}+c_{2/3} Q^{2/3} \tilde{R} \tilde{V}^{1/3} -\left[\frac{16}{9} \pi ^2 m_\sigma^2 N_f c_{2/3} c_{4/3} \nu^2  +\frac{9 (1-\gamma ) X_{00}^2 m_\pi^4 }{64 c_{4/3}^3 N_f^2}  + \cO\left( (1-\gamma )^2\right)\right] \tilde{V} \log Q +\cO\left(Q^0 \right)\ ,
     \end{split}
 \end{equation}
\begin{itemize}
    \item $\mathbf{V_{\Delta_\cO = 2}(\sigma)}$
     \end{itemize}
     \begin{equation}
\label{GSEnc2}
\small
    \begin{split}
            E^{\gamma \ll 1}&=\ \frac{c_{4/3} Q^{4/3}}{\tilde{V}^{1/3}}+Q^{2/3}\tilde{V}^{1/3}\left\{c_{2/3}\tilde{R}-\frac{1}{2}c_{2/3}m^2_\sigma-\frac{X_{00}^2}{4\pi^2 N_f^3 c_{4/3}^4}\bigg(\frac{9 m_\pi^2}{32\nu}\bigg)^2\left[1-\gamma\left(\frac{2}{3}\log Q-\frac{X_{10}}{X_{00}} \right. \right. \right. \\& \left. \left. \left. -\log\left(\frac{32 N_f\nu^2 \pi^2c_{4/3} \tilde{V}^{2/3}}{3}\right) + \cO\left( \gamma^2\right)\right)\right]\right\}+\left\{\frac{\gamma}{3\pi^2 N_f^4 c_{4/3}^5}\bigg(\frac{9 m_\pi^2}{32\nu}\bigg)^2\Bigg[\frac{5}{8\pi^2 c_{4/3}^4 N_f^2}\bigg(\frac{9 m_\pi^2}{32\nu}\bigg)^2 X_{00}^4 \right. \\& \left. -c_{2/3}\tilde{R} N_f X_{00}^2+\frac{9 X_{00} X_{01}}{32 c_{4/3}}\Bigg] + \cO\left( \gamma^2\right) \right\}\tilde{V} \log Q  +\cO\left(Q^0 \right)\ \qq{and}\\     
     E^{1-\gamma\ll 1}&=\frac{c_{4/3} }{\tilde{V}^{1/3}} Q^{4/3}+c_{2/3}Q^{2/3} \tilde{R} \tilde{V}^{1/3}-\frac{m_{\sigma}^2 c_{2/3} Q^{2/3} \tilde{V}^{1/3}}{2}-\left[\frac{9 (1-\gamma)  m_{\pi }^4 X_{00}^2 }{64  N_{f}^2 c_{4/3}^3}  + \cO\left( (1-\gamma )^2\right)\right]\tilde{V}\log Q+\cO\left(Q^0 \right)\ ,
     \end{split}
 \end{equation}
 \begin{itemize}
    \item $\mathbf{V_{\Delta_\cO \to 4}(\sigma)}$
     \end{itemize}
     \begin{equation}
\label{GSEnc3}
\small
    \begin{split}
           E^{\gamma \ll 1}&=\ \frac{c_{4/3} Q^{4/3}}{\tilde{V}^{1/3}}\left[1 +m^2_\sigma\frac{c_{2/3} }{256 \pi ^2 \nu ^2 N_f c_{4/3}^2}\left(4\log \left(\frac{3 \sqrt{\frac{3}{2}} Q}{128 \pi ^3 c_{4/3}^{3/2} \nu ^3 \tilde{V} N_f^{3/2}}\right)-3\right)+\order{m^4_\sigma}\right]\\&
 +c_{2/3} Q^{2/3}\tilde{R}\tilde{V}^{1/3}\Bigg\{1-\frac{X_{00}^2}{4\pi^2 N_f^3 c_{2/3} c_{4/3}^4 \tilde{R}}\bigg(\frac{9 m_\pi^2}{32\nu}\bigg)^2\Bigg[1-\gamma\left(\frac{2}{3}\log Q-\frac{X_{10}}{X_{00}}-\log\left(\frac{32 N_f\nu^2 \pi^2c_{4/3} \tilde{V}^{2/3}}{3}\right)\right)\Bigg]\\&-\frac{m_{\sigma}^2}{(8 c_{4/3}\pi \nu)^2 N_f}
 c_{2/3} 
 \log \left(\frac{3 \sqrt{\frac{3}{2}} Q}{128 \pi ^3 c_{4/3}^{3/2} \nu ^3 \tilde{V} N_f^{3/2}}\right)+\order{m^4_\sigma,m_\sigma^2 \gamma ,\gamma^2 }\Bigg\} +\tilde{V} \log Q \Bigg\{\frac{\gamma}{3\pi^2 N_f^4 c_{4/3}^5}\bigg(\frac{9 m_\pi^2}{32\nu}\bigg)^2  \\& \times \Bigg[\frac{5}{8\pi^2 c_{4/3}^4 N_f^2}\bigg(\frac{9 m_\pi^2}{32\nu}\bigg)^2 X_{00}^4-c_{2/3}\tilde{R} N_f X_{00}^2+\frac{9 X_{00} X_{01}}{32 c_{4/3}}\Bigg]+\frac{14 c_{2/3}^3 m_{\sigma }^2  \tilde{R}^2}{(32 \pi \nu)^2 N_f c_{4/3}^3}
 \Bigg\} 
 +\cO\left(Q^0 \right)\ \\          E^{1-\gamma\ll 1}&=\frac{c_{4/3} Q^{4/3} }{\tilde{V}^{1/3}} \left[1+\frac{ m_{\sigma}^2 c_{2/3}}{256 \pi ^2 N_{f} c_{4/3}^2 \nu^2}\left(4 \log \left(\frac{3 \sqrt{\frac{3}{2}} Q}{128 \pi ^3 c_{4/3}^{3/2} \nu ^3 \tilde{V} N_f^{3/2}}\right)-3\right) + \cO\left( m_\sigma^4\right)\right]\\& + c_{2/3}  Q^{2/3} \tilde{R} \tilde{V}^{1/3} \left[ 1 -\frac{  c_{2/3} m_{\sigma}^2}{(8 c_{4/3}\pi \nu)^2 N_f}
 \log \left(\frac{3 \sqrt{\frac{3}{2}} Q}{128 \pi ^3 c_{4/3}^{3/2} \nu ^3 \tilde{V} N_f^{3/2}}\right)  + \cO\left( m_\sigma^4\right)\right] \\& + \tilde{V}\log Q\left\{\frac{14 c_{2/3}^2 m_{\sigma }^2 \tilde{R}^2}{(32 \pi \nu)^2 N_f c_{4/3}^3}
 - \frac{9 (1-\gamma)  m_{\pi }^4 X_{00}^2 }{64  N_{f}^2 c_{4/3}^3}  + \cO \left(m_\sigma^4, m_\sigma^2 (1-\gamma),  (1-\gamma)^2  \right)\right\}+\cO\left(Q^0 \right)\ ,
     \end{split}
 \end{equation}
     
where we introduced
\begin{equation} \label{coeffic}
\small 
    c_{4/3}=\frac{3}{8}\left(\frac{\Lambda^2}{\pi N_f \nu^2}\right)^{2/3}, \quad c_{2/3}=\frac{1}{4 f^2}\left(\frac{\pi^2}{N_f \nu^2 \Lambda^4}\right)^{1/3}, \quad \tilde{R}=\frac{R}{6} \,,  \quad \tilde{V}=\frac{V}{2\pi^2} \,,  \quad \Lambda^4 \equiv \begin{cases}
        \Lambda_0^4 + \frac{m_\sigma^2}{16 f^2} \quad \Delta_\cO \to 0 \\  \Lambda_0^4 + \frac{m_\sigma^2}{8 f^2} \quad \ \Delta_\cO = 2 \\
         \Lambda_0^4  \qquad \qquad \Delta_\cO \to  4
    \end{cases}
\end{equation}
and we double-expanded $X$ first in $\gamma$ and then also in $1/Q$ as follows
\begin{align}
\label{expansion}
& X=X_0 + X_1  \gamma+ \cO\left(\gamma^2\right) \,, \qquad & X_k = X_{k0} + \frac{X_{k1}}{Q^{2/3}} + \cO\left(Q^{-4/3}\right) \,, \qquad &\text{for } \gamma \ll 1  \nonumber \\ &  X=X_0 + X_1  (1-\gamma)+ \cO\left((1-\gamma)^2\right) \,, \qquad & X_k = X_{k0} + \frac{X_{k1}}{Q^{4/3}} + \cO\left(Q^{-2}\right) \,, \qquad & \text{for } 1-\gamma \ll 1\ . 
\end{align} 
Within the same double expansion we will solve for $\bar \theta$. In particular, we will solve the equation of motion, that at the leading order in this expansion, yields $\bar \theta_{00}(\alpha_i)=0$. Henceforth, the latter is interpreted as double-expanding the $\alpha_i$. With a slight abuse of notation we use the same name for the coefficients of the expansions around $\gamma = 0$ and $1-\gamma=0$. Interestingly, in the latter case there is no term of order $\cO\left(Q^{-2/3} \right)$. 

For $\theta = 0$ eq.\eqref{GSEnc} reproduces the results in \cite{Orlando:2020yii} \footnote{Our expression corrects a few misprints in eqs.$(23)$ and $(24)$ of \cite{Orlando:2020yii}. We will give the corrected expression for $\theta=0$ in Sec.\ref{GSEaxion}}. When the theory is conformal (i.e. $m_\pi=m_\sigma=0$), the $\theta$-dependence disappears and the scaling dimension depends only on dimensionless coefficients as
\be \label{confdim}
\Delta_Q =\tilde{V}^{1/3}  E_Q= c_{4/3} Q^{4/3} +c_{2/3} Q^{2/3} + \cO\left(Q^0 \right) \,,
\ee
in agreement with the general form of the large-charge expansion for generic non-supersymmetric CFTs \cite{Gaume:2020bmp}. As first pointed out in \cite{Orlando:2019skh}, the deviations from conformality lead to corrections depending on the background geometry and involve the appearance of logarithms of $Q$. 

From eq.\eqref{GSEnc}, we observe that $\theta$ does not affect the leading order in the large-charge expansion. Moreover, we have that the $\theta$-vacuum is determined by $X_{00}$ which can be obtained by solving $\bar \theta_{00}(\alpha_i)=0$ in terms of the $\alpha_i$. In particular, $X_{00} = \sum_i^{N_f} \cos \alpha_{i00}$ (with the obvious notation) where the $\alpha_{i00}$s are the solutions found in \cite{Bersini:2022jhs} for the $\alpha_i$ variables at the leading order in $\frac{m_\pi^2}{a}$ and we will refer to such results in the text. In particular, in the $1-\gamma\ll 1$ case, the minimum energy is achieved when $X_{00}^2$ is maximized due to the minus sign in front of the $\log Q$ term in the energy \eqref{GSEnc}. It follows that the problem of maximizing $X_{00}$ in terms of the $\alpha_{i00}$ is equivalent to the problem studied in \cite{Bersini:2022jhs} of maximizing $X^2$ in terms of the $\alpha_i$ which determines the vacuum structure of the superfluid phase of theory in absence of the dilaton. In other words, the results of \cite{Bersini:2022jhs} apply here by replacing $X \to X_{00}$ and $\alpha_i \to \alpha_{i00}$ and the $\theta$-dependence of the vacuum in the presence of the dilaton is identical to the case without it. 

Interestingly, when $\gamma\ll 1$ the situation is more subtle since the leading $\theta$-dependent contribution is of the form $X_{00}^2 \left(\frac23\gamma \log Q-1\right)$. Therefore, at fixed $\gamma$ there exists a critical charge below/above which the $\theta$-vacuum has to maximize/minimize $X_{00}^2$. In other words, there is a phase transition in the large-charge regime as $Q$ crosses 
\be
Q_c = e^{\frac{3}{2 \gamma }} \ .
\ee
 Therefore, for $Q < Q_c$ the $\theta$-vacuum will be analogous to the $1-\gamma\ll 1$ case. We expect this to be the natural limit here as well given that for very charges larger than $Q_c$ we should start exploring the UV asymptotically free fixed point rather than the IR interacting one. 
 
As we showed in \cite{Bersini:2022jhs}, the $\alpha_{i00}$ assume the following expressions that solve eq.\eqref{eqalphai}:
\begin{align} \label{Solgen2}
    \alpha_{i00}=\begin{cases}
  \pi-  \alpha ,\qquad  & i=1,\dots,n \\
  \alpha ,\qquad & i=n+1,\dots,N_f
    \end{cases} 
\end{align}
where $\alpha$ is given by
\begin{equation} \label{Solgen}
    \alpha=\frac{\theta+ (2k-n)\pi}{(N_{f}-2n)},\quad k=0,\dots, N_f-2n-1 \,, \quad n=
    0,..., \left[\frac{N_f-1}{2}\right] \,.
\end{equation}
As discussed in \cite{Bersini:2022jhs}, $X_{00}^2$ is always maximized by solutions with $n=0$. For even $N_f$ the ground state solutions have $\alpha_{i00} = \frac{\theta}{N_f}$ in the interval $\theta \in [0, \pi]$ and $\alpha_{i00} = \frac{\theta-2 \pi}{N_f}$ for $\theta \in [\pi, 2\pi]$. The energies stemming from the two solutions cross at $\theta = \pi$ where for $N_f > 2$ we have spontaneous $CP$ symmetry breaking. For odd $N_f$, the solutions that maximize $X_{00}^2$ are $\alpha_{i00} = \frac{\theta}{N_f}$ (for $\theta \in [0, \frac{\pi}{2}]$), $\alpha_{i00} =  \frac{\theta -\pi }{N_f}+\pi$ (for $\theta \in [ \frac{\pi}{2}, \frac{3\pi}{2}]$) , $\alpha_{i00} = \frac{\theta-2 \pi}{N_f}$ (for $\theta \in [\frac{3\pi}{2}, 2\pi]$). In particular, this entails that for odd $N_f$, $CP$ is conserved at $\theta=\pi$, being the vacuum non-degenerate, while we have two first-order phase transitions at $\theta=\frac{\pi}{2}$ and $\theta=\frac{3\pi}{2}$. 

We conclude this section by determining the $\theta$-dependent parameters entering the ground state energy for $\gamma \ll 1$ case when $Q<Q_c$ where $X_{00} = N_f \cos \alpha$. We obtain 
\begin{align}
    X_{00}&= N_f \cos{\left(\frac{\theta+2k \pi}{N_f}\right)}\\
    X_{01}&= \frac{9 m_\pi^4 \sin ^2\left(\frac{\theta+2k \pi}{N_f}\right) \cos \left(\frac{\theta+2k \pi}{N_f}\right)}{8\  a \ c_{4/3}^2 }\\
     \bar \theta_{01}&=\frac{m_\pi^2 X_{00} \sin (\frac{\theta+2\pi k}{N_f})}{a N_f}\\
       \bar \theta_{11}&=\frac{3 m_\pi^2 \sin \left(\frac{2 (\theta +2 \pi  k)}{N_f}\right) \log \left(\frac{8192 \pi ^2 c_{4/3}^3 N_f^3 v^6}{27 Q^2}\right)}{32\  a \ c_{4/3}^2} \\
    X_{10} &= X_{11}= \bar \theta_{00}= \bar \theta_{10} =0  
\end{align}


where the value of $k$ relevant for a given value of $\theta$ has to be taken in accordance with the previous discussion.

\subsection{Ground state energy with the axion}
\label{GSEaxion}

In this section, we focus on the theory in the presence of the axion field described by eq.\eqref{lagaxion} and consider the ansatz \eqref{eq:ansatsvacuum} for the ground state of $\Sigma$ and \cite{DiVecchia:2013swa}
\be \label{realeq12}
\langle N \rangle = e^{-i\delta/a_{PQ}} 
\ee
for the axion field. This leads us to study the following Lagrangian
\begin{equation}
\label{lagcontheta}
\begin{split}
\small
    \mathcal{L}\left[\Sigma_0,\sigma_0, \delta \right]&=-a \nu ^2 e^{-4 f \sigma_0 } (\Bar{\theta}-\delta )^2-e^{-4 f \sigma_0 } \left(\Lambda ^4-\frac{m_\sigma^2}{16 f^2}\right)-V(\sigma_0)+\\&+2 \nu ^2 m_\pi^2 X \cos \varphi\  e^{-f \sigma_0  y}+2 \mu ^2 \nu ^2 N_f e^{-2 f \sigma_0 } \sin ^2\varphi\ -\frac{R e^{-2 f \sigma_0 }}{12 f^2}
\end{split}
\end{equation}
and the corresponding EOMs:
\begin{align}
     \frac{\delta \mathcal{L}}{\delta \alpha}=\frac{\delta \mathcal{L}}{\delta \varphi}=\frac{\delta \mathcal{L}}{\delta \sigma_{0}}=\frac{\delta \mathcal{L}}{\delta \delta}=0, \quad \frac{\delta \mathcal{L}}{\delta \mu}=\frac{Q}{V}\ .
\end{align}
The main difference with the axion-less case is expressed in the equations for the $\alpha_i$ and $\delta$:
\begin{align}
    a e^{-4 f \sigma_0 } (\bar \theta-\delta)-2m_{\pi}^2 \sin \alpha_i \cos \varphi  e^{-f \sigma_0  y}&=0 \,, \qquad i=1, .., N_f \\
     \delta -\bar \theta&=0\ .
\end{align}
These equations can be solved as $\delta = \theta$ and $\alpha_i=0$. Thus the $\theta$-dependence disappears from the ground state energy and there is no $CP$ violation since $(\bar \theta-\delta)=0$. In fact, by construction the axion realizes the Peccei-Quinn solution of the strong $CP$ problem \cite{peccei1977cp,peccei1977constraints} and, as a consequence, the classical solutions are equivalent to the ones at $\theta=0$ which are obtained by setting $X_{00} = N_f$ and $X_{01}=\bar \theta_{01} =0$ in eqs.\eqref{GSEnc}, \eqref{GSEnc2}, \eqref{GSEnc3}. For instance, in the $V_{\Delta_\cO =0}(\sigma)$ case, one obtains the results in \cite{Orlando:2020yii}
\begin{equation}
 \label{CCWy3}
 \small
 \begin{split}
     E^{\gamma \ll 1}&=\ \frac{c_{4/3} Q^{4/3}}{\tilde{V}^{1/3}}+Q^{2/3}\tilde{V}^{1/3}\left\{c_{2/3}\tilde{R}-\frac{1}{4\pi^2 N_f c_{4/3}^4}\bigg(\frac{9 m_\pi^2}{32\nu}\bigg)^2\left[1-\gamma\left(\frac{2}{3}\log Q-\log\left(\frac{32 N_f\nu^2 \pi^2c_{4/3} \tilde{V}^{2/3}}{3}\right)\right)\right]\right\}+\\&-\tilde{V} \log (Q) \left\{\frac{16\pi^2}{9}N_f c_{2/3}c_{4/3}\nu^2m_\sigma^2-\frac{\gamma}{3\pi^2 N_f^2 c_{4/3}^5}\bigg(\frac{9 m_\pi^2}{32\nu}\bigg)^2\left[\frac{5}{8\pi^2 c_{4/3}^4}\bigg(\frac{9 m_\pi^2}{32\nu}\bigg)^2-c_{2/3}\tilde{R} N_f\right]\right\}
 \end{split}
 \end{equation}
for the case of $\gamma\ll 1$, while for $1-\gamma\ll 1$ we have
 \begin{equation}
 \label{CCWy2}
 \small
     E^{1-\gamma\ll 1}=\frac{c_{4/3} Q^{4/3}}{\tilde{V}^{1/3}}+c_{2/3} Q^{2/3} \tilde{R} \tilde{V}^{1/3} -\frac{9 (1-\gamma ) m_\pi^4 \tilde{V} \log Q}{64 c_{4/3}^3}-\frac{16}{9} \pi ^2 m_\sigma^2 N_f c_{2/3} c_{4/3} \nu^2 \tilde{V} \log Q\ .
 \end{equation}
What differs from \cite{Orlando:2020yii} is the presence in the spectrum of a light axion which we must take into account when considering the second-order expansion. 
\section{Second-order expansion in the near-conformal regime}
\label{SOENC}
The spectrum of the theory in absence of the dilaton field has been detailed  in \cite{Bersini:2022jhs}. There, we observed that the theory is characterized by the following symmetry-breaking pattern  
\begin{equation}
    SU(2N_f)\times U(1)_A \overset{2N_f^2-N_f}{\rightsquigarrow}Sp(2N_f)\longrightarrow SU(N_f)_V\times U(1)_B \overset{\frac{N_f^2-N_f}{2}}{\rightsquigarrow} Sp(N_f)_V
\end{equation}
where $\rightsquigarrow$ and $\longrightarrow$ denote, respectively, spontaneous and explicit breaking. In fact, the axial symmetry is explicitly broken by the anomaly and therefore the would-be Goldstone boson is massive. The further explicit breaking is due to the introduction of a baryon charge while the last spontaneous breaking is the superfluid transition. In the absence of the dilaton, the spectrum of light modes is composed of $\frac{1}{2}N_f(N_f-1)$ massless Goldstones with speed $v_G=1$ that parameterizes the coset $\frac{SU(N_f)}{Sp(N_f)}$. They arrange themselves in the antisymmetric representation of the unbroken $Sp(N_f)$ plus a singlet which we denote as $\pi_0$. In addition, we find $\frac{1}{2}N_f(N_f-1)-1$ pseudo-Goldstone with mass $\frac{m_\pi^2 X}{\mu N_f}$ plus a pseudo-Goldstone mode stemming from the would-be spontaneous breaking of $U(1)_A$ which we call the $S$ (singlet) mode. As mentioned above, the $U(1)_A$ symmetry is  quantum mechanically anomalous, and therefore the latter mode acquires a mass contribution proportional to the scale of the anomaly, $\sqrt{a}$.\\ The rest of the spectrum is given by gapped modes with a mass of order $\mu$ which we will not consider here since they decouple from the large-charge dynamics.
It is interesting to analyze how the spectrum changes when (near)conformal dynamics is realized through the dilaton dressing. In particular, conformal invariance dictates the existence of a massless mode with speed $v_G = \frac{1}{\sqrt{d-1}}= \frac{1}{\sqrt{3}}$ \cite{Orlando:2019skh}.\\ 
As we shall see, the latter arises from the mixing between the singlet mode $\pi_0$ with the dilaton that acts as its "radial mode" and changes its speed from $v_G =1$ to $v_G= \frac{1}{\sqrt{3}}$. In general, the infrared dynamics of this mode can be described by using a conformally invariant four derivative action of the type $\mathcal{L}_{NLSM} = k_4 (\partial_\mu \chi \partial^\mu \chi)^2$. In turn, the latter can be seen as the heavy-dilaton limit of the two-derivative action $\mathcal{L}_{2} = \nu^2 (\partial_\mu \chi \partial^\mu \chi)$ after we dress it with the dilaton \cite{Hellerman:2015nra,Son:2002zn,Monin:2016jmo,Orlando:2019skh}.\\ Having in mind the hierarchy of scales $m_\pi, m_\sigma \ll \sqrt{a} \leq \mu \ll 4 \pi \nu$, we focus on the spectrum of light modes i.e. the modes whose mass is smaller than the chemical potential in the large-charge limit. To this end, we start by considering the modes with mass $\frac{m_\pi^2 X}{\mu N_f}$. These cannot mix with the dilaton and, therefore, their dispersion relation in the near-conformal phase can be obtained by generalizing the non-conformal expression of \cite{Bersini:2022jhs} to include the expectation value of the dilaton as
\begin{equation}
\label{padressed}
    \omega_1^2=k^2 -\frac{m_\pi^4 X^2 e^{-2 f \sigma_0 (y-2)}}{ \mu ^2 N_f^2}\ ,
\end{equation}
which matches eq.(59) of \cite{Bersini:2022jhs} when $f = 0$.\\ For $y=2$ this mode has squared mass $\frac{m_\pi^4 X^2}{ \mu ^2 N_f^2}$, whereas in the  limit $y=3$ we have

\begin{equation}
  \omega_1^2=k^2 -   \frac{27 m_\pi^4 X_{00}^2}{512 \pi ^2 \nu ^2 N_f^3 c_{4/3}^3} + \cO\left(Q^{-2/3} \right)\ .
\end{equation}
\newline
To study the remaining light modes, we expand around the vacuum solution as follows
\begin{align}
\label{perttheta}
   \Sigma=e^{i \Omega} \Sigma_{0}e^{i \Omega^{t}}
\end{align}
where $\Sigma_{0}$ is the classical solution \eqref{eq:ansatsvacuum}
while the fluctuations are organized in the matrix $\Omega$ as 
\begin{align}
 \Omega=   \left(\begin{array}{cc}
\pi & 0 \\
0 & -\pi^{t}
\end{array}\right)+\tilde{\beta} S\left(\begin{array}{ll}
\mathbbm{1}_{N_f} & 0 \\
0 & \mathbbm{1}_{N_f}
\end{array}\right), \quad\tilde{\beta} \equiv \frac{1}{\sqrt{2 N_f}}
\end{align}
 here $\pi=\sum_{a=0}^{\text{dim}\frac{U(N_f)}{Sp(N_f)}} \pi^{a}T_{a}$ belongs to the algebra of the coset space $\frac{U(N_f)}{Sp(N_f)}$. The normalization condition on the generators is $\left\langle  T_{a}T_{b}\right\rangle=\frac{\delta_{ab}}{2}$. Using the results contained in App.\ref{appB}, we find
\begin{align}
    \left\langle\partial_{\mu} \Sigma \partial^{\mu} \Sigma^{\dagger}\right\rangle&=4\sin^2\varphi\ \partial_{\mu}\pi^{a}\partial^{\mu}\pi^{a}+8N_{f} \tilde{\beta}^2 \partial_{\mu}S\partial^{\mu}S\\
\left\langle B \Sigma \partial_{0} \Sigma^{\dagger}\right\rangle&=-2i\sqrt{
    2 N_{f}}  \sin ^{2}\varphi \partial_{0} \pi^{0}\\
\left\langle M\Sigma+M^{\dagger}\Sigma^{\dagger}\right\rangle&=2N_{f}\cos\varphi\left(X \cos \left( \sqrt{\frac{2}{N_f}} S\right)+Z \sin \left(\sqrt{\frac{2}{N_f}} S\right)\right)\\
\left\langle \log \Sigma-\log \Sigma^{\dagger}\right\rangle&=8i N_{f}\tilde{\beta}S-\sum_i^{N_f}\alpha_i\ ,
\end{align}
where we defined $Z \equiv \sum_{i=1}^{N_f} \sin \alpha_i$.
Finally, by expanding the dilaton field around its background solution as $\sigma\rightarrow\sigma_{0}+\hat{\sigma}(t,\mathbf{x})$, we obtain the following quadratic Lagrangian
\begin{align}
\label{ldivkin}
\frac{\mathcal{L}}{4\nu^2\sin^2\varphi\ e^{-2\sigma_{0}f}}=\left(\begin{array}{lll}
\pi^{0} & \hat{\sigma} &S
\end{array}\right)D^{-1}\left(\begin{array}{l}
\pi^{0} \\
\hat{\sigma} \\
S
\end{array}\right)+\sum_{a=1}^{\text{dim}\left({\tiny\Yvcentermath1 \yng(1,1)
}\right)}\partial^{\mu}\pi^{a}\partial_{\mu}\pi^{a}
\end{align}
with the inverse propagator $D^{-1}$  defined as
\begin{align}
\label{d-1theta}
  D^{-1}=\left(\begin{array}{ccc}
\omega^{2}-k^{2} & i\omega\mu f \sqrt{2N_{f}} & 0 \\
-i\omega\mu f \sqrt{2N_{f}}& \frac{\omega^{2}-k^{2}}{8\nu^2\sin^2\varphi}-M_{\sigma}^{2} & \frac{ 1}{2} I_{\hat{\sigma}s} \\
0 &\frac{1}{2} I_{\hat{\sigma}s} &\frac{\omega^{2}-k^{2}}{\sin^2\varphi}-M_{s}^{2}
\end{array}\right)  \,, \qquad   I_{\hat{\sigma}S} =\frac{\sqrt{2} f \mu ^2 m_\pi^4 \sqrt{N_f} X y Z}{m_\pi^4 X^2-\mu ^4 N_f^2 e^{2 f \sigma_0 (y-2)}} 
\end{align} 
where the Lagrangian masses for the dilaton field  are given by

\begin{itemize}
    \item $\mathbf{V_{\Delta_\cO \to 0}(\sigma)}$
     \end{itemize}
\begin{equation}
\small
    M_{\sigma}^2=-\frac{f^2 \mu ^2 N_f e^{-6 f \sigma_0} \left(\nu ^2 m_\pi^4 X^2 \left(y^2-2\right) e^{6 f \sigma_0}+2 \mu ^4 \nu ^2 N_f^2 e^{2 f \sigma_0 (y+1)}-4 \Lambda ^4 \mu ^2 N_f e^{2 f \sigma_0 y}\right)}{2 \nu ^2 \left(\mu ^4 N_f^2 e^{2 f \sigma_0 (y-2)}-m_\pi^4 X^2\right)}
\end{equation}
\begin{itemize}
    \item $\mathbf{V_{\Delta_\cO = 2}(\sigma)}$
     \end{itemize}

\begin{equation}
\small
    M_{\sigma}^2=-\frac{\mu ^2 e^{-f \sigma_0 y} \left(-8 f^2 e^{f \sigma_0 y} \left(2 \mu ^2 \left(a v^2 {\bar \theta}^2+\Lambda ^4\right)+m_\pi^4 N_f v^2 X^2\right)+4 f^2 m_\pi^4 N_f v^2 X^2 y^2 e^{3 f \sigma_0}+\mu ^2 e^{f \sigma_0 (y+2)} \left(8 f^2 \mu ^2 N_f v^2+m_\sigma^2\right)\right)}{8 v^2 \left(\mu ^4 e^{2 f \sigma_0}-m_\pi^4 X^2\right)}
\end{equation}

\begin{itemize}
    \item $\mathbf{V_{\Delta_\cO \to 4}(\sigma)}$
    
     \end{itemize}

\begin{equation}
    \small
    M_{\sigma}^2=-\frac{\mu ^2 \left(4 f^2 \left(4 \mu ^2 \left(a v^2 {\bar \theta}^2+\Lambda ^4\right)+m_\pi^4 N_f v^2 X^2 \left(2-y^2 e^{-f \sigma_0 (y-3)}\right)-2 \mu ^4 N_f v^2 e^{2 f \sigma_0}\right)-4 f \mu ^2 m_\sigma^2 \sigma_0+\mu ^2 m_\sigma^2\right)}{8 v^2 \left(m_\pi^4 X^2-\mu ^4 e^{2 f \sigma_0}\right)}
\end{equation}
while the mass of the $S$ mode is the same for all three potentials and reads     
\normalsize
\begin{equation} \label{massofS}
    M_{S}^2=\frac{a \mu ^4 N_f^3 e^{2 f \sigma_0 (y-1)}+2 \mu ^2 m_\pi^4 X^2 e^{4 f \sigma_0}}{2 \mu ^4 N_f^2 e^{2 f \sigma_0 y}-2 m_\pi^4 X^2 e^{4 f \sigma_0}}\ .
\end{equation}
The $\pi^a$ represents Goldstone modes with sound speed $v_G$=1 transforming in the antisymmetric representation of $Sp(N_f)$ analogous to the ones found in absence of the dilaton field in \cite{Bersini:2022jhs}. 
The $\pi_0$ field corresponds to the Goldstone mode of the coset $\frac{SU(N_f)}{Sp(N_f)}$ that transforms as a singlet of $Sp(N_f)$. For this reason, it mixes with the dilaton and the $S$ according to eq.\eqref{d-1theta}.
At this point, one can investigate whether eq.\eqref{massofS} is in agreement with the Witten-Veneziano relation \cite{witten1979current,veneziano1979u}. The latter constrains the mass squared of the $S$-particle to be proportional to the number of flavours and the topological susceptibility in the limit $m_\pi \to 0$. 
Indeed, this agrees with the following limit of eq.\eqref{massofS} \footnote{Our definition of $M_{S}$ differs by a factor of $2$ from the usual conventions.}
\begin{equation} \label{WV}
    \lim_{\sigma_0\to 0,\ m_\pi\to 0} \eqref{massofS} \implies M^2_{S}=\frac{a N_f}{2}\ .
\end{equation}
Additionally, for $\sigma_0\to 0$ we have
\begin{equation}
    \lim_{\sigma_0\to 0} \eqref{massofS} \implies M^2_{S}=\frac{a \mu ^4 N_f^3+2 \mu ^2 m_\pi^4 X^2}{2 \mu ^4 N_f^2-2 m_\pi^4 X^2}\ ,
\end{equation}
in agreement with eq.(64) of \cite{Bersini:2022jhs}.
The remaining dispersion relations are obtained by solving the equation $ \text{det}(D^{-1})=0$. As a result, we have one massless degree of freedom and two gapped modes.
In what follows we provide explicit expressions for the dispersion relations at the two extrema of the dimension of the fermion condensate $\gamma=0$ and $\gamma=1$ in the limit of small momenta and within the large-charge expansion. For the sake of simplicity, we consider only the case $\Delta_\cO \to 0$. For the massless mode we obtain
\begin{align}
\small
 \gamma=0 &:\quad\omega_2=\sqrt{\frac{9 f^2 m_\pi^8 X_0^2 Z_0^2 e^{-4 f \sigma_0}-2 \mu ^4 M^2_\sigma M^2_{S} N_f^3 \sin^4\varphi\ }{9 f^2 m_\pi^8 X_0^2 Z_0^2 e^{-4 f \sigma_0}-2 \mu ^4 M^2_{S} N_f^3 \sin^4\varphi\  \left(2 f^2 \mu ^2 N_f+M^2_\sigma\right)}}\ k + \cO\left(k^2 \right)\\
\gamma= 1&:\quad\omega_2=\sqrt{\frac{2 f^2 m_\pi^8 N_f X_0^2 Z_0^2-\mu ^4 M^2_\sigma M^2_{S} \sin^4\varphi\ }{2 f^2 N_f \left(m_\pi^8 X_0^2 Z_0^2-\mu ^6 M^2_{S} \sin^4\varphi\ \right)-\mu ^4 M^2_\sigma M^2_{S} \sin^4\varphi\ }}\ k + \cO\left(k^2 \right)
\end{align}
where we expanded $Z$ in $\gamma$ and $1-\gamma$ as for $X$ in eq.\eqref{expansion} and kept only the leading term. In the large-charge limit, the above reduces to
\begin{align}
\small
\label{wy3}
 \gamma=0 &:\quad\omega_2=k \left[\frac{1}{\sqrt{3}}+\ \frac{\sqrt{3}\ X_{00}^2 }{\left( 2\pi ^2\right)^{2/3} c_{4/3}^5 N_f^3}\left(\frac{9 m_\pi^2 }{128 \pi  \nu }\right)^2\  \left(\frac{V}{Q}\right)^{2/3}+\dots\right] + \cO\left(k^2 \right)\\
\label{wy2}
\gamma=1&:\quad\omega_2=k \left[\frac{1}{\sqrt{3}}+1\ \left(\frac{ 2^{5/3} c_{2/3} \nu ^2 m_\sigma^2 }{3 \sqrt{3} \pi ^{2/3}}+\frac{9 \sqrt{3} m_\pi^4 X_{00}^2}{128 \sqrt[3]{2} \pi ^{8/3} c_{4/3}^4 N_f^2}\right) \left(\frac{V}{Q}\right)^{4/3}+\dots\right] +\cO\left(k^2 \right)\ .
\end{align}
These results agree with those in \cite{Orlando:2020yii} for $\theta=0$ i.e. $X_{00}=N_f$ (see Sec.\ref{questaqua} for further details). At the conformal point $m_\sigma=m_\pi=0$, eqs.\eqref{wy3} and \eqref{wy2} describe the expected conformal phonon whose speed approaches the value $v_G= \frac{1}{\sqrt{3}}$ as $Q\to \infty$ \cite{Orlando:2019skh}. 

For the massive modes described by eq.\eqref{d-1theta}, we write the dispersion relations as an expansion for small momenta
\begin{align}\label{massive1}
     \omega_3&=M_3 + v_3 k^2 + \dots\\\label{massive2}
    \omega_4&= M_4 + v_4 k^2 + \dots
\end{align}
In the $\gamma=0$ case, we have
\footnotesize
\begin{align}
    M_{3,4}&=\frac{1}{\sqrt{2}}\sqrt{8 \nu ^2 \sin^2\varphi \left(2 f^2 \mu ^2 N_f+M^2_\sigma\right)+M^2_{S} \sin^2\varphi \pm\frac{\sqrt{144 f^2 \nu ^2 m_\pi^8 X_0^2 Z_0^2 e^{-4 f \sigma_0}+\mu ^4 N_f^3 \sin^4\varphi \left(M^2_{S}-8 \nu ^2 \left(2 f^2 \mu ^2 N_f+M^2_\sigma\right)\right)^2}}{\mu ^2 N_f^{3/2}}}\\
    v_{3,4}&=\frac{\mu ^4 N_f^3 e^{4 f \sigma_0} \left(-2  M_{3,4}^2 \sin^2\varphi \left(8 \nu ^2 \left(f^2 \mu ^2 N_f+M^2_\sigma\right)+M^2_{S}\right)+3  M_{3,4}^4+8 \nu ^2 M^2_\sigma M^2_{S} \sin^4\varphi\right)-36 f^2 \nu ^2 m_\pi^8 X_0^2 Z_0^2}{2  M_{3,4} \mu ^4 N_f^3 e^{4 f \sigma_0} \left(8 \nu ^2 \sin^2\varphi \left(M^2_{S} \sin^2\varphi-2  M_{3,4}^2\right) \left(2 f^2 \mu ^2 N_f+M^2_\sigma\right)-2 l M_{3,4}^2 M^2_{S} \sin^2\varphi+3  M_{3,4}^4\right)-72  M_{3,4} f^2 \nu ^2 m_\pi^8 X_0^2 Z_0^2}
\end{align}\normalsize
which in the large-charge limit yields \small
\begin{align}
   \omega_3&=\frac{8\ 2^{5/6} \pi ^{2/3} f c_{4/3} \nu  \sqrt{N_f}}{\sqrt{3} }\left(\frac{Q}{V}\right)^{1/3}+\frac{ \left(640 \pi ^2 k^2 N_f^2 c_{4/3}^3-567 f^2 m_\pi^4 X_{00}^2\right)}{2048\ 2^{5/6} \sqrt{3} \pi ^{8/3} f c_{4/3}^4 \nu  N_f^{5/2} }\left(\frac{V}{Q}\right)^{1/3}+\dots\\
   \omega_4&=\frac{\sqrt{3 a}  }{4\ 2^{2/3} \pi^{1/3}  \sqrt{c_{4/3}} \nu  }\left(\frac{Q}{V}\right)^{1/3}+\left[\frac{9\sqrt{3} m_\pi^4  X_{00}^2 \left(9 a+256 \pi ^2 c_{4/3}^3 \nu ^2\right)}{4096\ 2^{10/3} \pi ^{11/3} \sqrt{a} c_{4/3}^{11/2} \nu ^3 N_f^3 }+\frac{2^{5/3} \pi^{1/3} k^2 \sqrt{c_{4/3}} \nu  }{\sqrt{3} \sqrt{a} }\right]\left(\frac{V}{Q}\right)^{1/3}+\dots
\end{align}\normalsize
\\For the case $\gamma=1$, the mass and speed of the massive modes are given by \footnotesize
\begin{align}
    M_{3,4}&=\frac{\sin\varphi}{\sqrt{2}} \sqrt{ 8 \nu ^2 \left(2 f^2 \mu ^2 N_f+M^2_\sigma\right)+M^2_{S} \pm \sin^2\varphi \sqrt{\frac{64 f^2 \nu ^2 m_\pi^8 X_0^2 Z_0^2+\mu ^4 N_f^3 \sin^4\varphi \left(M^2_{S}-8 \nu ^2 \left(2 f^2 \mu ^2 N_f+M^2_\sigma\right)\right)^2}{\mu ^4 N_f^3 \sin^8\varphi}}}\\
    v_{3,4}&=\frac{\mu ^4 \left(-2  M_{3,4}^2 \sin^2\varphi \left(8 \nu ^2 \left(f^2 \mu ^2 N_f+M^2_\sigma\right)+M^2_{S}\right)+3 M_{3,4}^4+8 \nu ^2 M^2_\sigma M^2_{S} \sin^4\varphi\right)-16 f^2 \nu ^2 m_\pi^8 N_f X_0^2 Z_0^2}{\text{Den}}\\
     \text{Den}&=2 M_{3,4}^3 \mu ^4 \left(3 M_{3,4}^2-2 M^2_{S} \sin^2\varphi\right)-16 M_{3,4} \nu ^2 \big[2 f^2 N_f \left(\mu ^6 \sin^2\varphi \left(2 M_{3,4}^2-M^2_{S} \sin^2\varphi\right) m_\pi^8 X_0^2 Z_0^2\right)\nonumber\\
    &+\mu ^4 M^2_\sigma \sin^2\varphi \left(2 M_{3,4}^2-M^2_{S} \sin^2\varphi\right)\big]
\end{align}\normalsize
whose large-charge limit gives \small
\begin{align}
    \omega_3&=\frac{8\ 2^{5/6} \pi ^{2/3} f \nu  \sqrt{N_f} c_{4/3} }{\sqrt{3} }\left(\frac{Q}{V}\right)^{1/3}+\frac{5 k^2 }{16\ 2^{5/6} \sqrt{3} \pi ^{2/3} f \nu  \sqrt{N_f} c_{4/3} }\left(\frac{V}{Q}\right)^{1/3}+\dots\\
    \omega_4&=\frac{\sqrt{3 a} }{4\ 2^{2/3} \pi^{1/3} \sqrt{c_{4/3}} \nu  }\left(\frac{Q}{V}\right)^{1/3}+\frac{ 2^{5/3} \pi^{1/3} k^2 \sqrt{c_{4/3}} \nu  }{\sqrt{3} \sqrt{a} }\left(\frac{V}{Q}\right)^{1/3}+\dots\quad
\end{align} \normalsize
Notice that the $\gamma=1$ case leads to enhanced suppression of the $\theta$-dependence in the large-charge limit compared to the $\gamma=0$ case as already observed at the classical level.

\subsection{Second-order expansion with the axion}
\label{questaqua}
Similarly to the previous section, we parametrize the fluctuations of the $\Sigma$ field as in \eqref{perttheta} and we take into account the fluctuations of the axion field that we write as \cite{DiVecchia:2013swa}
\begin{equation}
    N(x)= e^{i\hat{a} (x)/\nu_{PQ} }\ \expval{N}
\end{equation}
with $\expval{N}$ given by eq.\eqref{realeq12}. Therefore, \eqref{lagaxion} gives another term to consider with respect to the previous section which is
\begin{equation}
    -\frac{i}{2}a_{PQ}(\log N - \log N^\dagger)=\frac{a_{PQ}}{\nu_{PQ}}\hat{a}-\delta\ .
\end{equation}
As a consequence, we again have modes with dispersion relation \eqref{padressed} with $X_{00}=N_f$, while the remaining light fluctuations are described by the Lagrangian below
\begin{align}
\label{lagaxionfluc}
\mathcal{L}&=4 \nu^{2} \sin ^{2} \varphi  \partial_{\mu} \pi^{a} \partial^{\mu}\pi^{a}  e^{-2 \sigma f}+
4 \nu^2 \partial_{\mu} S \partial^{\mu} S e^{-2 \sigma f}+
8 \sqrt{2 N} \mu\nu^2 \sin ^{2} \varphi\  \partial_{0} \pi^{0} e^{-2 \sigma f}+2 \nu^{2} \mu^{2} N_{f} \sin ^{2} \varphi e^{-2 \sigma f}+\nonumber\\
&+
2N_{f}^2\nu^2m^2_\pi\cos\varphi \cos \left( \sqrt{\frac{2}{N_f}} S\right) e^{-y \sigma f}-a \nu^{2}\left(\Bar{\theta}+\sqrt{2N_{f}}S+\frac{a_{PQ}}{2\nu_{PQ}}\hat{a}-\delta\right)^{2} e^{-2 \sigma f}\nonumber\\
&-\Lambda_{0}^{4} e^{-4 \sigma f}-\frac{R}{12 f^{2}} e^{-2 \sigma f}+\big(\partial_\mu \hat{a}\ \partial^\mu \hat{a}\big) e^{-2 \sigma f}+\frac{1}{2}\big(\partial_\mu\sigma\partial^\mu\sigma\big) e^{-2 \sigma f}-V(\sigma)\ ,
\end{align}
from which we obtain the normalised quadratic Lagrangian as
\begin{align}
\frac{\mathcal{L}}{4\nu^2\sin^2\varphi\ e^{-2\sigma_{0}f}}=\left(\begin{array}{llll}
\pi^{0} & \hat{\sigma} &S &\hat{a}
\end{array}\right)D^{-1}\left(\begin{array}{l}
\pi^{0} \\
\hat{\sigma} \\
S\\
\hat{a}
\end{array}\right)+\sum_{a=1}^{\text{dim}\left(\tiny\Yvcentermath1 \yng(1,1)\right)}\partial^{\mu}\pi^{a}\partial_{\mu}\pi^{a}
\end{align}
where the inverse propagator $D^{-1}$ is given by
\begin{align}
\label{d-1axion}
\small
D^{-1} =  \left(\begin{array}{cccc}
\omega^{2}-k^{2} & i\omega\mu f \sqrt{2N_{f}} & 0 & 0 \\
-i\omega\mu f \sqrt{2N_{f}}& \frac{\omega^{2}-k^{2}}{8\nu^2\sin^2\varphi}-M_{\sigma}^{2} & 0 & 0\\
0 & 0 &\frac{\omega^{2}-k^{2}}{\sin^2\varphi}-M_{S}^{2} &\frac{I_{S \hat{a}}}{2}\\
0 & 0 & \frac{I_{S \hat{a}}}{2} & \frac{\omega^{2}-k^{2}}{4\nu^2\sin^2\varphi}-M_{\hat{a}}^{2} 
\end{array}\right)\ .  
\end{align}
The Lagrangian masses appearing in $D^{-1}$ are

\begin{itemize}
    \item $\mathbf{V_{\Delta_\cO \to 0}(\sigma)}$
     \end{itemize}
\begin{equation}
    M^2_\sigma=-\frac{f^2 \mu ^2  e^{-2 f \sigma_0 (y-2)} \left(\nu ^2 m_\pi^4 N_f \left(y^2-2\right) e^{2 f \sigma_0 (y-2)}+2 \mu ^4 \nu ^2 N_f e^{4 f \sigma_0 (y-2)}-4 \Lambda ^4 \mu ^2  e^{2 f \sigma_0 (2 y-5)}\right)}{2 \nu ^2 \left(\mu ^4  e^{2 f \sigma_0 (y-2)}-m_\pi^4 \right)} ,
\end{equation}
\begin{itemize}
    \item $\mathbf{V_{\Delta_\cO = 2}(\sigma)}$
     \end{itemize}

\begin{equation}
\small
    M_{\sigma}^2=-\frac{4 f^2 \mu ^2 \nu ^2 m_{\pi}^4 N_{f} \left(y^2-2\right)+\mu ^4 e^{2 f \sigma_{0} (y-2)} \left(8 f^2 \mu ^2 \nu ^2 N_{f}+m_{\sigma}^2\right)-16 f^2 \Lambda ^4 \mu ^4 e^{2 f \sigma_{0} (y-3)}}{8 \nu ^2 \left(\mu ^4 e^{2 f \sigma_{0} (y-2)}-m_{\pi}^4\right)}, 
\end{equation}

\begin{itemize}
    \item $\mathbf{V_{\Delta_\cO \to 4}(\sigma)}$
     \end{itemize}

\begin{equation}
    \small
    M_{\sigma}^2=-\frac{2 f^2 \mu ^2 \nu ^2 m_{\pi }^4 N_{f} \left(y^2-2\right)+\mu ^4 e^{2 f \sigma_{0} (y-3)} \left(m_{\sigma}^2 (2 f \sigma_{0}-1)-8 f^2 \Lambda ^4\right)+4 f^2 \mu ^6 \nu ^2 N_{f} e^{2 f \sigma_{0} (y-2)}}{4 \nu ^2 \left(\mu ^4 e^{2 f \sigma_{0} (y-2)}-m_{\pi}^4\right)}
\end{equation}
for the dilaton and
\begin{equation}
     M^2_{S}=\frac{a \mu ^4 N_f e^{2 f \sigma_0 (y-1)}+2 \mu ^2 m_\pi^4  e^{4 f \sigma_0}}{2 \mu ^4  e^{2 f \sigma_0 y}-2 m_\pi^4  e^{4 f \sigma_0}},
\end{equation}
\begin{equation}
     M^2_{\hat{a}}=\frac{a \ a_{PQ}^2 e^{-2 f \sigma_0}}{16 \nu_{PQ}^2 \left(1-\frac{m_\pi^4  e^{-2 f \sigma_0 (y-2)}}{\mu ^4 }\right)}\ 
\end{equation}
for the remaining modes. The interaction term between the $S$-particle and the fluctuation of the axion $\hat{a}$ is
\begin{equation}
    I_{S \hat{a}}=-\frac{\sqrt{N_f}\ a \mu ^4  a_{PQ} e^{2 f \sigma_0 (y-3)}}{2 \sqrt{2} \nu_{PQ} \left(\mu ^4  e^{2 f \sigma_0 (y-2)}-m_\pi^4 \right)}\ .
\end{equation}
The inverse propagator \eqref{d-1axion} takes the form of a block matrix. The upper left $2 \times 2$ block represents the mixing between $\pi_0$ and the $\hat{\sigma}$ in the absence of the $\theta$-angle. Diagonalizing it, we find both a massless and a gapped mode with dispersion relations given by
\begin{equation}
    \omega_{5.6} = \sqrt{k^2+4 \nu ^2 \sin ^2 \phi  \left(2 f^2 \mu ^2 N_f+M_{\sigma }^2 \pm\sqrt{\frac{f^2 k^2 \mu ^2 N_f \csc ^2\phi }{\nu ^2}+\left(2 f^2 \mu ^2 N_f+M_{\sigma }^2\right){}^2}\right)}
\end{equation}

The spectrum in the $\Delta_\cO \to 0$ case has been previously studied in \cite{Orlando:2020yii}. In that case, the explicit expression of the dispersion relation of the massless mode in the two extreme cases $\gamma= 0$ and $\gamma = 1$ reads
\begin{equation}
    \small
    \gamma=0:\quad \omega_5=\frac{k}{\sqrt{3}}+k\ \frac{\sqrt{3} }{\left( 2\pi ^2\right)^{2/3} c_{4/3}^5 N_f}\left(\frac{9 m_\pi^2 }{128 \pi  \nu }\right)^2\  \left(\frac{V}{Q}\right)^{2/3}+\dots
\end{equation}
\begin{equation}
    \small
    \gamma=1:\quad \omega_5=\frac{k}{\sqrt{3}} +k\ \left[\frac{2\ 2^{2/3} c_{2/3} \nu ^2 m_\sigma^2 N_f}{3 \sqrt{3} \pi ^{2/3}}+\frac{9 \sqrt{3} m_\pi^4 }{256 \sqrt[3]{2} \pi ^{8/3} c_{4/3}^4}\right]\left(\frac{V}{Q}\right)^{4/3}+\dots
\end{equation}
The above results correct a typo in eqs.$(53)$ and $(54)$ of \cite{Orlando:2020yii}. The gapped mode arising from the mixing between $\pi_0$ and the dilaton has a mass of order $\cO\left(\mu \right)$. The lower right $2 \times 2$ block describes the mixing between the $S$ mode and the axion $\hat{a}$ and by diagonalizing it we obtain the dispersion relations of the propagating modes which read
\begin{align}
    \omega_{7,8}= \frac{1}{2} \sqrt{4 k^2+2 \sin^2\varphi \left(4 \nu ^2 M^2_{\hat{a}}+M^2_{S}\right) \pm \frac{e^{-2 f \sigma_0}}{\nu_{PQ}}\sqrt{ 2 a^2 \nu ^2 N_f a_{PQ}^2+4 \nu_{PQ}^2 \sin^4\varphi \ e^{4 f \sigma_0} \left(M^2_{S}-4 \nu ^2 M^2_{\hat{a}}\right)^2}}
\end{align}
For $f=0$ the above reproduces the dispersion relations in absence of the dilaton, which were computed in \cite{Bersini:2022jhs}. Finally, we provide explicit results for these dispersion relations in the $\Delta_\cO \to 0$ case. For $\gamma= 0$ and in the large-charge expansion, we have  
\begin{align}
    \omega_7&=\frac{ \sqrt{\frac{3 a \tilde{\nu}}{c_{4/3} N_f}}}{8 2^{1/6} \pi^{1/3} \nu_{PQ} \nu  }\left(\frac{Q}{V}\right)^{1/3}+\frac{3^5\ }{16^4 }\frac{\sqrt{a} m_\pi^4 \sqrt{\tilde{\nu}_{PQ}}}{2^{2/3} \sqrt{3} \pi ^{11/3} \nu_{PQ} c_{4/3}^{11/2} \nu ^3 N_f^{3/2}}\left(\frac{V}{Q}\right)^{1/3}+\nonumber\\
    &+ \frac{\sqrt{N_f}\nu_{PQ} \left[\nu_{PQ}^2 \left(512 \pi ^2 k^2 c_{4/3}^3 \nu ^2 N_f+27 m_\pi^4\right)+256 \pi ^2 k^2 c_{4/3}^3 \nu ^4 a_{PQ}^2\right]}{\sqrt{3 a} \pi ^{5/3} c_{4/3}^{5/2} \nu   \tilde{\nu}_{PQ}^{3/2}}\left(\frac{V}{Q}\right)^{1/3}+\dots\\
    \omega_8&=\frac{\sqrt{512 \pi ^2 k^2 c_{4/3}^3 N_f \tilde{\nu}+27 m_\pi^4 a_{PQ}^2 }}{16 \pi  c_{4/3}^{3/2}\sqrt{2\tilde{\nu} N_f}}+\nonumber\\
    &+\frac{3^3 m_\pi^8 a_{PQ}^2 N_f^{5/2}  \left(27 a \tilde{\nu}^2-2048 \pi ^2 \nu_{PQ}^4 c_{4/3}^3 \nu ^2 N_f^2\right)}{16^3\cdot 8^2\cdot 2^{1/6} \pi ^{13/3} a c_{4/3}^{13/2} \nu ^2  \tilde{\nu}^{5/2} \sqrt{512 \pi ^2 k^2 c_{4/3}^3 N_f \tilde{\nu}+27 m_\pi^4 a_{PQ}^2 }}\left(\frac{V}{Q}\right)^{2/3}+\dots
\end{align}
while for $\gamma = 1$ we obtain
\begin{align}
    \omega_7&=\frac{\sqrt{3 \tilde{\nu}}}{8\cdot 2^{1/6}\nu_{PQ}\pi^{1/3}\nu\sqrt{N_f c_{4/3}}}\left(\frac{Q}{V}\right)^{1/3}+\frac{4\cdot 2^{1/6} \pi^{1/3} \nu_{PQ} k^2 \sqrt{c_{4/3}} \nu  \sqrt{N_f} }{\sqrt{3 a \tilde{\nu}}}\left(\frac{V}{Q}\right)^{1/3}+\dots\\
    \omega_8&=k+\frac{9 \nu ^2 m_\pi^4 a_{PQ}^2  }{32\ 2^{2/3} \pi ^{4/3} k c_{4/3}^2  \tilde{\nu}}\left(\frac{V}{Q}\right)^{2/3}+\dots
\end{align}
with $\tilde{\nu}\equiv 2\nu^2_{PQ}N_f+a_{PQ}^2 \nu^2$.

The upshot of this subsection is that the effect of eliminating the $\theta$-angle via the axion, in the large charge regime, is the introduction of a new light state affecting the spectrum and dynamics of the theory expressed in the new dispersion relations $\omega_7$ and $\omega_8$.

\subsection{Conformal dimension and vacuum energy of type I Goldstones}

As eq.\eqref{confdim} illustrates, in the conformal limit $m_\pi=m_\sigma=0$, the scaling dimension of the lowest-lying operator with baryon charge $Q$ takes the form predicted by the large-charge EFT \cite{Hellerman:2015nra, Monin:2016jmo}  
\be \label{confindustria}
\Delta_Q = k_{4/3} Q^{4/3} +k_{2/3} Q^{2/3} + k_0 \log Q + \cO\left(Q^0 \right) \,,
\ee
where the coefficients that appear should not be confused with $c_{4/3}$ and $c_{2/3}$ introduced in eq.\eqref{coeffic}. Instead the latter should be viewed as the leading order of a semiclassical expansion of the former, generated by integrating out the heavy degrees of freedom when building the large charge EFT. By inspection, we found out that the parameter controlling the leading quantum correction to the classical result \eqref{GSEnc} is 
\be 
\frac{c_{4/3}}{c_{2/3}^2} =\frac{6}{\pi^2} f^4 \Lambda^4 \,.
\ee
This can be traced back to the observations made in \cite{Orlando:2019skh}, where the authors mapped a dilaton-dressed $2$-derivatives action with $U(1)$ symmetry (analogous to the $\pi_0$, $\hat{\sigma}$ sub-sector of the present theory) to the familiar $\lambda \phi^4$ model with quartic coupling $\lambda = f^4 \Lambda^4$. In fact, it is known that within this model the Wilson coefficient of the large charge EFT can be computed as a semiclassical expansion in $\lambda$ by considering the double-scaling limit $Q \to \infty$, $\lambda \to 0$ with 't Hooft-like coupling $\lambda Q$ fixed \cite{Badel:2019oxl, Antipin:2020abu}. The computation of the leading quantum correction to $\Delta_Q$ is lengthy and beyond the scope of the present work. However, as observed in \cite{Cuomo:2020rgt}, the knowledge of the spectrum of gapless modes is enough to compute exactly the $k_0$ coefficient in eq.\eqref{confindustria}, which is related to the renormalization of their vacuum energy. In our case, the massless spectrum described by the large charge EFT is composed of the $\frac{1}{2} N_f (N_f-1)-1$ modes with dispersion relation $\omega_1$and the $\frac{1}{2} N_f (N_f-1)$ $\pi^a$ modes. In the conformal limit, these are $N_f^2 - N_f- 1$ type I Goldstone bosons of the spontaneous symmetry breaking (see \cite{Bersini:2022jhs} for details and transformation properties)
\be
SU(N_f)_L \times SU(N_f)_R \times U(1)_B \times U(1)_A \rightsquigarrow Sp(N_f)_L \times Sp(N_f)_R 
\ee
and all have sound speed $v_G=1$ with the exception of the conformal mode $\pi_0$ with $v_G=\frac{1}{\sqrt{3}}$. We do not include the singlet mode $S$ since it decouples due to the axial anomaly. Consider the action describing a type I Goldstone field $\chi$ at low energy
\begin{equation}
    S_G=\int_{\mathcal{M}}\ \dd t \dd \mathbf{x}\ \left(\frac{1}{2}(\partial_t\chi)^2+\frac{v^2_G}{2}\big(\grad{\chi}\big)^2\right)\ ,
\end{equation}
its one-loop vacuum energy reads
\begin{equation}
\label{flucdet}
    E_{\rm Casimir}=\frac{1}{2}Tr\{\log\big(-\partial^2_t-v^2_G \laplacian\big)\} = \frac{1}{4\pi}\int_{-\infty}^{\infty}\ \dd\omega \sum_{\mathbf{p}}\ \log\big(\omega^2+v^2_G E^2(\mathbf{p})\big) = \frac{v_G}{2} \sum_{\mathbf{p}} E(\mathbf{p})
\end{equation}
where $E(\mathbf{p})^2$ are the eigenvalues of the Laplacian on the sphere, i.e. $\laplacian f_{\mathbf{p}}(\mathbf{r})+ E^2(\mathbf{p}) f_{\mathbf{p}}(\mathbf{r})=0$. Being this contribution of order $Q^0$, in odd dimension is not renormalized by the classical vacuum energy and is, therefore, universal to all the CFT whose large-charge dynamics realize the same conformal superfluid phase \cite{Gaume:2020bmp, Hellerman:2015nra}. Conversely, in the case of an even number of spacetime dimensions, there is a universal $Q^0\log Q$ term arising from the renormalization of the energy. In fact, as shown in \cite{Cuomo:2020rgt}, in dimensional regularization $E_G$ has a pole for $d \to 4$ which is linked to a calculable logarithm of the charge with coefficient $-\frac{v_G}{48}$.
Hence, we obtain
\begin{equation}
   k_0=-\frac{1}{48}\left(\frac{1}{\sqrt{3}}+ (N_f+1)(N_f-2) \right) .
\end{equation}
This is a robust non-perturbative prediction of the large-charge approach which would be interesting to test via lattice simulations.

\section{Conclusions and outlook}
\label{Conclusions}

In this work we investigated the impact of the $\theta$-angle and axion physics in the large-charge limit of near-conformal symplectic gauge theories with fermion matter in the defining representation. The simplest symplectic gauge theory is $SU(2)$ with $N_f$ fermions in the fundamental representation. We modelled the underlying dynamics via an effective approach encapsulating the light degrees of freedom of the theory augmented by a dilaton state and an eta-like prime state that ensures the axial-anomaly variation of the underlying theory at the effective level. The large-charge approach is then employed to access non-perturbative information of the (near) conformal fixed baryon-charge sector of the theory.   Using the state-operator correspondence we  generalised  the corrections determined in \cite{Orlando:2019skh,Orlando:2020yii} to the large-charge quasi-anomalous dimension $\Delta$ as a function of the dilaton, fermion mass, and  background geometry to include the impact of the $\theta$ angle, axion physics, and dilaton potential yielding: 


  \subsubsection*{ $\mathbf{V_{\Delta_\cO \to 0}(\sigma)}$, $\gamma \ll1$} \small
  \begin{align} \frac{\Delta}{ \Delta^\ast} &=1 -\left(\frac{9 m_{\pi }^2}{32 \pi   \nu }\right)^2\frac{   1-\gamma  \log \left(\frac{3 \rho ^{2/3}}{16 (2 \pi^2)^{1/3}  c_{4/3} \nu ^2 N_f}\right)}{4 c_{4/3}^5 N_f}\ \textcolor{rossoCP3}{\cos ^2\left(\frac{\theta +2 \pi  k}{N_f}\right)} \left(\frac{1}{2 \pi ^2 \rho }\right)^{2/3}  \nonumber \\ &+\frac{\gamma}{c_{4/3}^6 N_f}   \textcolor{rossoCP3}{\cos ^2\left(\frac{\theta +2 \pi  k}{N_f}\right)} \left(\frac{27 m_{\pi }^4\  \textcolor{rossoCP3}{\sin ^2\left(\frac{\theta +2 \pi  k}{N_f}\right)}}{256\ a\ c_{4/3}^3 N_f^2}+\frac{5 \left(\frac{9 m_{\pi }^2}{64 \pi  \nu }\right)^2 \textcolor{rossoCP3}{\cos ^2\left(\frac{\theta +2 \pi  k}{N_f}\right)}}{6 c_{4/3}^4 N_f}-\frac{ c_{2/3}}{2} \left(\frac{2 \pi^2\rho}{ Q}\right)^{2/3}\right) \nonumber \\  & \times \left(\frac{9 m_{\pi }^2}{32 \pi  \nu }\right)^2\left(\frac{1}{2 \pi ^2 \rho }\right)^{4/3} \log Q \textcolor{cadmiumorange}{-\frac{16}{9} \pi ^2 c_{2/3} \nu ^2 N_f m_{\sigma }^2 \left(\frac{1}{2 \pi ^2 \rho }\right)^{4/3} \log Q} \end{align}
\normalsize
 \subsubsection*{ $\mathbf{V_{\Delta_\cO \to 0}(\sigma)}$, $(1-\gamma) \ll1$} \small \begin{equation} \frac{\Delta}{ \Delta^\ast} = 1- \frac{9 m_{\pi }^4 }{64 c_{4/3}^4} (1-\gamma )\  \textcolor{rossoCP3}{\cos ^2\left(\frac{\theta +2 \pi  k}{N_f}\right)}\left(\frac{1}{2 \pi ^2 \rho }\right)^{4/3} \log Q\textcolor{cadmiumorange}{-\frac{16}{9} \pi ^2 c_{2/3} \nu ^2 N_f m_{\sigma }^2 \left(\frac{1}{2 \pi ^2 \rho }\right)^{4/3} \log Q}  \ .\end{equation}
 \normalsize
   \subsubsection*{ $\mathbf{V_{\Delta_\cO = 2}(\sigma)}$, $\gamma \ll1$} \small
  \begin{align} \frac{\Delta}{ \Delta^\ast} &=1 -\left(\frac{9 m_{\pi }^2}{32 \pi   \nu }\right)^2\frac{   1-\gamma  \log \left(\frac{3 \rho ^{2/3}}{16 (2 \pi^2)^{1/3}  c_{4/3} \nu ^2 N_f}\right)}{4 c_{4/3}^5 N_f}\ \textcolor{rossoCP3}{\cos ^2\left(\frac{\theta +2 \pi  k}{N_f}\right)} \left(\frac{1}{2 \pi ^2 \rho }\right)^{2/3}  \textcolor{cadmiumorange}{-\frac{c_{2/3} m_{\sigma }^2}{2 c_{4/3}}  \left(\frac{1}{2 \pi ^2 \rho }\right)^{2/3} }\nonumber \\ &+\frac{\gamma}{c_{4/3}^6 N_f}   \textcolor{rossoCP3}{\cos ^2\left(\frac{\theta +2 \pi  k}{N_f}\right)} \left(\frac{27 m_{\pi }^4\  \textcolor{rossoCP3}{\sin ^2\left(\frac{\theta +2 \pi  k}{N_f}\right)}}{256\ a\ c_{4/3}^3 N_f^2}+\frac{5 \left(\frac{9 m_{\pi }^2}{64 \pi  \nu }\right)^2 \textcolor{rossoCP3}{\cos ^2\left(\frac{\theta +2 \pi  k}{N_f}\right)}}{6 c_{4/3}^4 N_f}-\frac{ c_{2/3}}{2} \left(\frac{2 \pi^2 \rho}{ Q}\right)^{2/3}\right) \nonumber \\  & \times \left(\frac{9 m_{\pi }^2}{32 \pi  \nu }\right)^2\left(\frac{1}{2 \pi ^2 \rho }\right)^{4/3} \log Q \end{align}
\normalsize

 \subsubsection*{ $\mathbf{V_{\Delta_\cO = 2}(\sigma)}$, $(1-\gamma) \ll1$} \small \begin{equation} \frac{\Delta}{ \Delta^\ast} = 1\textcolor{cadmiumorange}{-\frac{ c_{2/3} m_{\sigma }^2}{2 c_{4/3}} \left(\frac{1}{2 \pi ^2 \rho }\right)^{2/3}}-  \frac{9 m_{\pi }^4 }{64 c_{4/3}^4} (1-\gamma )\  \textcolor{rossoCP3}{\cos ^2\left(\frac{\theta +2 \pi  k}{N_f}\right)}\left(\frac{1}{2 \pi ^2 \rho }\right)^{4/3} \log Q \ .\end{equation}
 \normalsize
   \subsubsection*{ $\mathbf{V_{\Delta_\cO \to 4}(\sigma)}$, $\gamma \ll1$} \small
  \begin{align} \frac{\Delta}{ \Delta^\ast} &=1\textcolor{cadmiumorange}{-\frac{c_{2/3} m_{\sigma }^2 }{256 \pi ^2 c_{4/3}^2 \nu ^2 N_f} \left(3-4 \log \left(\frac{3 \sqrt{\frac{3}{2}} \rho }{64 \pi  c_{4/3}^{3/2} \nu ^3 N_f^{3/2}}\right)\right) +\frac{c_{2/3}^2 m_{\sigma }^2 }{256 \pi ^2 c_{4/3}^3 \nu ^2 N_f} \left(3-8 \log \left(\frac{3 \sqrt{\frac{3}{2}} \rho }{64 \pi  c_{4/3}^{3/2} \nu ^3 N_f^{3/2}}\right)\right) \frac{1}{ Q^{2/3}}} \nonumber \\& -\left(\frac{9 m_{\pi }^2}{32 \pi   \nu }\right)^2\frac{   1-\gamma  \log \left(\frac{3 \rho ^{2/3}}{16 (2 \pi^2)^{1/3}  c_{4/3} \nu ^2 N_f}\right)}{4 c_{4/3}^5 N_f}\ \textcolor{rossoCP3}{\cos ^2\left(\frac{\theta +2 \pi  k}{N_f}\right)} \left(\frac{1}{2 \pi ^2 \rho }\right)^{2/3} 
  \end{align}
\normalsize

 \subsubsection*{ $\mathbf{V_{\Delta_\cO \to 4}(\sigma)}$, $(1-\gamma) \ll1$} \small \begin{align} \frac{\Delta}{ \Delta^\ast} &=1\textcolor{cadmiumorange}{-\frac{c_{2/3} m_{\sigma }^2 }{256 \pi ^2 c_{4/3}^2 \nu ^2 N_f}\left(3-4 \log \left(\frac{3 \sqrt{\frac{3}{2}} \rho }{64 \pi  c_{4/3}^{3/2} \nu ^3 N_f^{3/2}}\right)\right)+ \frac{c_{2/3}^2 m_{\sigma }^2 }{256 \pi ^2 c_{4/3}^3 \nu ^2 N_f} \left(3-8 \log \left(\frac{3 \sqrt{\frac{3}{2}} \rho }{64 \pi  c_{4/3}^{3/2} \nu ^3 N_f^{3/2}}\right)\right) \frac{1}{ Q^{2/3}}} 
 \end{align}
 \normalsize
\vskip 1cm
Here $\Delta^\ast$ is the conformal dimension at the fixed point at  leading order in the semiclassical expansion and $\rho \equiv Q/V$ is the charge density.

We summarise below our main results: 
\begin{itemize}
\item[a)]{
We determined the effects coming from different values of $\Delta_\cO$ of the underlying operator responsible for conformal breaking. These are encoded in the dilaton potential and their results are summarised in the orange part of the deformed conformal dimensions. The first observation is that all corrections are universally proportional to the dilaton squared mass  while the dependence on $\Delta_\cO$ is encoded in both the coefficient and in the scaling dependence on the charge (density) of the $m^2_\sigma$ term. For  $\Delta_\cO \to 0$  the leading conformal breaking term scales  with $\nu^2 \rho^{-4/3} \log Q$ \cite{Orlando:2019skh}  while for  $\Delta_\cO = 2$ and $\Delta_\cO = 4$  we have respectively, $ \rho^{-2/3}$  and $\nu^{-2} \log (\rho/\nu^3)$ at fixed but large charge density.}

\item[b)]{Conformal breaking due to the introduction of the $\pi$ mass, to the chiral and conformal lagrangian order that we are considering, yields an additive contribution to the deformed anomalous dimensions with respect to the dilaton one. This contribution is independent of the value of $\Delta_\cO$ and it is universally proportional to the fourth power of the pion mass (i.e. quadratic in the underlying fermion masses). Its contribution depends, however, on the value of the anomalous dimension of the fermion-antifermion condensate operator controlled by $\gamma$. In the $\Delta_\cO \to 0$  case the zero $\theta$-angle expression was determined in \cite{Orlando:2020yii} while here we provide, for the first time, its $\theta$-angle dependence. These terms are highlighted in red. For the $\Delta_\cO \to 4$ we also have these terms but they are subleading at large charge density with respect to the dilaton contribution. Nevertheless, they are naturally there when the dilaton mass vanishes.}

\end{itemize}

\section*{Acknowledgments}

M. T. was supported by Agencia Nacional de Investigación y Desarrollo (ANID) grant 72210390. The work of J.B. is supported by the World Premier International Research Center Initiative (WPI Initiative), MEXT, Japan.

\newpage
\begin{center}
 \LARGE\textbf{\textsc{APPENDIX}}  
\end{center}

\begin{appendices}
\renewcommand\thesubsection{\thesection.\arabic{subsection} }
 \renewcommand{\theequation}{\thesection.\arabic{equation}}
\section{Details of the second-order expansion}
\label{appB}
\setcounter{equation}{0}
In this Appendix, we provide details on the derivation of the dispersion relations of the fluctuations that we parametrize as 
\begin{align}
   \Sigma=e^{i \Omega} \Sigma_{0}e^{i \Omega^{t}}
\end{align}
where $\Sigma_{0}$ corresponds to homogeneous ground state \eqref{eq:ansatsvacuum}
while $\Omega$ has the form
\begin{align}
 \Omega=   \left(\begin{array}{cc}
\pi & 0 \\
0 & -\pi^{t}
\end{array}\right)+\tilde{\beta} S\left(\begin{array}{ll}
\mathbbm{1}_{N_f} & 0 \\
0 & \mathbbm{1}_{N_f}
\end{array}\right) = \nu + \tilde{\beta} S \ \mathbbm{1}_{2 N_f}, \quad\tilde{\beta}=\frac{1}{\sqrt{2 N_f}}
\end{align}
 here $\pi=\sum_{a=0}^{\text{dim}} \pi^{a}T_{a}$ belongs to the algebra of the coset space $\frac{U(N_{f})}{Sp(N_{f})}$. The normalization condition on the generators is $\left\langle  T_{a}T_{b}\right\rangle=\frac{\delta_{ab}}{2}$. 
Writing $\Sigma=e^{2 i \tilde{\beta} S}U(\alpha_{i})\overline{\Sigma}$, we have that $\overline{\Sigma}$ can be written as \small
\begin{align}
\overline{\Sigma}&=e^{i\nu}\Sigma_c e^{i\nu^t}=\left(\begin{array}{cc}
e^{ i\pi} & 0 \nonumber\\
0 & e^{-i\pi^t}\end{array}\right)\left( \left(\begin{array}{cc}
0 & \mathbbm{1}_{N_f} \nonumber \\
-\mathbbm{1}_{N_f} & 0
\end{array}\right) \cos \varphi+\left(\begin{array}{cc}
\sigma_{2}\otimes \mathbbm{1}_{N_{f}/2} & 0 \\
0 & \sigma_{2}\otimes \mathbbm{1}_{N_{f}/2} 
\end{array}\right)\sin \varphi\right)\left(\begin{array}{cc}
e^{ i\pi} & 0 \\
0 & e^{-i\pi^t}\end{array}\right)\\
&=\left(\begin{array}{ll}
e^{i \pi} & 0 \\
0 & e^{-i\pi^{t}}
\end{array}\right)\left(\begin{array}{cc}
0 & \mathbbm{1}_{N_f} \\
-\mathbbm{1}_{N_f} & 0
\end{array}\right)\left(\begin{array}{cc}
e^{i \pi^{t}} & 0 \\
0 & e^{-i\pi}
\end{array}\right) \cos \varphi\nonumber\\
&+\left(\begin{array}{cc}
e^{i \pi} & 0 \\
0 & e^{-i \pi^{t}}
\end{array}\right)\left(\begin{array}{cc}
\sigma_{2} \otimes \mathbbm{1}_{N_{f} / 2} & 0 \\
0 & \sigma_{2} \otimes \mathbbm{1}_{N_{f} / 2}
\end{array}\right)\left(\begin{array}{ll}
e^{i \pi^{t}} & 0 \\ 
0 & e^{-i\pi}
\end{array}\right)\sin\varphi \nonumber\\
&=\left(\begin{array}{cc}
0 & \mathbbm{1}_{N_f} \\
-\mathbbm{1}_{N_f} & 0
\end{array}\right) \cos \varphi+e^{2i\nu}\mathbbm{1}_{2}\otimes\sigma_{2} \otimes \mathbbm{1}_{N_{f} / 2}\sin\varphi.\label{sig}
\end{align} \normalsize
Using this expression is easy to show that (for simplicity of notation, henceforth we indicate the trace operation with the brackets)
\begin{align}
    \partial_{\mu} \overline{\Sigma}&=2 i  e^{2i\nu}\partial_{\mu} \nu \mathbbm{1}_{2}\otimes\sigma_{2} \otimes \mathbbm{1}_{N_{f} / 2}\sin\varphi,\\
    \partial_{\mu} \overline{\Sigma}^{\dagger}&=-2 i    \mathbbm{1}_{2}\otimes\sigma_{2} \otimes \mathbbm{1}_{N_{f} / 2}\partial_{\mu}\nu e^{-2i\nu}\sin\varphi,\\
    \overline{\Sigma}\partial_{0}\overline{\Sigma}^{\dagger}&=\left(\begin{array}{cc}
0 & -\sigma_{2}\otimes \mathbbm{1}_{N_{f}}\partial_{0}\pi^{t} e^{-2i \pi^{t}} \\
-\sigma_{2}\otimes \mathbbm{1}_{N_{f}}\partial_{0}\pi e^{-2i \pi}& 0
\end{array}\right)\sin\varphi\cos \varphi-2i\partial_{0}\nu\sin^2\varphi
\end{align}
which implies $\left\langle\partial_{\mu} \overline{\Sigma} \partial^{\mu} \overline{\Sigma}^{\dagger}\right\rangle=8 \sin ^{2}(\varphi)\left\langle\partial_{\mu} \pi \partial^{\mu} \pi\right\rangle$ and $\left\langle B \overline{\Sigma} \partial_{0} \overline{\Sigma}^{\dagger}\right\rangle=-2 i \sin ^{2}(\varphi)\left\langle\partial_{0} \pi\right\rangle$ where $B=\frac{1}{2}\sigma_{3}\otimes \mathbbm{1}_{N_{f}}$. Additionally we have
\begin{align}
\left\langle M\overline{\Sigma}+M^{\dagger}\overline{\Sigma}^{\dagger}\right\rangle&=4 N_{f}\cos\varphi\\
    \left\langle B \overline{\Sigma} B^{t} \overline{\Sigma}^{\dagger}\right\rangle&=\left\langle B e^{i \nu} \Sigma_{0}e^{i \nu^{t}} B^{t} e^{-i \nu^{t}}  \Sigma_{0}^{\dagger}e^{-i\nu}\right\rangle \nonumber \\
    &=\left\langle B \Sigma_{0} B^{t} \Sigma_{0}^{\dagger}\right\rangle
\end{align}
since $[B,\nu]=0$. 

Making use of the above results, we obtain that the matrix $\Sigma$ satisfies
\begin{align}
\partial_{\mu}\Sigma&=e^{2i  \tilde{\beta}S}U(\alpha_{i})\left(2 i \tilde{\beta}  \partial_{\mu}  S\overline{\Sigma}+ \partial_{\mu}\overline{\Sigma}\right),\\
\partial_{\mu}\Sigma^{\dagger}&=e^{-2i  \tilde{\beta}S}U(\alpha_{i})\left(-2 i \tilde{\beta}  \partial_{\mu} S\overline{\Sigma}^{\dagger}+ \partial_{\mu}\overline{\Sigma}^{\dagger}\right)\\
\Sigma\partial_{\mu}\Sigma^{\dagger}&=-2i \tilde{\beta}\partial_{\mu}S+\overline{\Sigma}\partial_{\mu}\overline{\Sigma}^{\dagger}\ .
\end{align}
From which we can obtain the relevant traces as
\begin{align}
    \left\langle\partial_{\mu} \Sigma \partial^{\mu} \Sigma^{\dagger}\right\rangle&=\left\langle\partial_{\mu} \overline{\Sigma} \partial^{\mu} \overline{\Sigma}^{\dagger}\right\rangle+4\tilde{\beta}^2\left\langle\partial_{\mu} S \partial^{\mu} S\right\rangle+2i \tilde{\beta}\partial^{\mu}S\left\langle\overline{\Sigma}\partial_{\mu}\overline{\Sigma}^{\dagger}\right\rangle\nonumber\\
    &=8 \sin ^{2}\varphi\left\langle\partial_{\mu} \pi \partial^{\mu} \pi\right\rangle+8N_{f} \tilde{\beta}^2 \partial_{\mu}S\partial^{\mu}S\nonumber\\
    &=4\sin^2\varphi\ \partial_{\mu}\pi^{a}\partial^{\mu}\pi^{a}+8N_{f} \tilde{\beta}^2 \partial_{\mu}S\partial^{\mu}S
\\
\left\langle B \Sigma \partial_{0} \Sigma^{\dagger}\right\rangle&=\left\langle B\overline{\Sigma}\partial_{0}\overline{\Sigma}^{\dagger}\right\rangle\nonumber\\
    &=-2i  \sin ^{2}\varphi\left\langle\partial_{0} \pi\right\rangle \nonumber\\&=-2i\sqrt{
    2 N_{f}}  \sin ^{2}\varphi\ \partial_{0} \pi^{0}\\
\left\langle M\Sigma+M^{\dagger}\Sigma^{\dagger}\right\rangle&=\left\langle e^{2i\tilde{\beta}S}MU(\alpha_{i})\overline{\Sigma}+e^{-2i \tilde{\beta}S}M^{\dagger}U(\alpha_{i})^{\dagger}\overline{\Sigma}^{\dagger}\right\rangle\nonumber\\
&=2N_{f}\cos\varphi\left(X \cos \left( \sqrt{\frac{2}{N_f}} S\right)+Z \sin \left(\sqrt{\frac{2}{N_f}} S\right)\right)\\
\left\langle \log \Sigma-\log \Sigma^{\dagger}\right\rangle&=8i N_{f}\tilde{\beta}S-\sum_i^{N_f}\alpha_i\ .
\end{align}
Finally, by plugging these results into the Lagrangian \eqref{lagdressed}, we arrive at
\begin{align}
\label{lagthetafluc}
\mathcal{L}&=4 \nu^{2} \sin ^{2} \varphi  \partial_{\mu} \pi^{a} \partial^{\mu}\pi^{a}  e^{-2 \sigma f}+
4 \nu^2 \partial_{\mu} S \partial^{\mu} S e^{-2 \sigma f}+
8 \sqrt{2 N} \mu\nu^2 \sin ^{2} \varphi \partial_{0} \pi^{0} e^{-2 \sigma f}\nonumber\\
&+2 \nu^{2} \mu^{2} N_{f} \sin ^{2} \varphi e^{-2 \sigma f}+
4 N_{f} \nu^{2} m_\pi^{2} \cos \varphi \left[X \cos \left(\sqrt{2} \sqrt{\frac{1}{N_f}} S\right)+Z \sin \left(\sqrt{2} \sqrt{\frac{1}{N_f}} S\right)\right] e^{-y \sigma f}+\nonumber\\
&-a v^{2}(\Bar{\theta}+\sqrt{2N_{f}}S)^{2} e^{-2 \sigma f}-\Lambda_{0}^{4} e^{-4 \sigma f}-\frac{R}{12 f^{2}} e^{-2 \sigma f}+\frac{1}{2}\big(\partial_\mu\sigma\partial^\mu\sigma\big) e^{-2 \sigma f}-V(\sigma)\ .
\end{align}
By expanding the above to the quadratic order in the fluctuations of the dilaton field, we obtain the quadratic Lagrangian \eqref{ldivkin} considered in the main text.

\end{appendices}
\printbibliography
\end{document}